\documentclass[10pt,twocolumn]{article}

\usepackage[UTF8,scheme=plain]{ctex}

\usepackage{newtxtext,newtxmath}

\usepackage{graphicx}
\usepackage{float}\usepackage{placeins}%

\usepackage[paperwidth=8.25in,paperheight=10.5in,left=0.5in,right=0.5in,top=0.55in,bottom=0.62in,footskip=0.3in]{geometry}
\makeatletter
\renewcommand\normalsize{\@setfontsize\normalsize{8.5}{10.5}%
  \abovedisplayskip 6pt plus 2pt minus 2pt\belowdisplayskip \abovedisplayskip
  \abovedisplayshortskip 2pt plus 2pt\belowdisplayshortskip 4pt plus 2pt minus 1pt
  \let\@listi\@listI}
\renewcommand\small{\@setfontsize\small{7.5}{9}%
  \abovedisplayskip 4pt plus 2pt minus 2pt\belowdisplayskip \abovedisplayskip}
\renewcommand\footnotesize{\@setfontsize\footnotesize{7}{8.5}}
\renewcommand\scriptsize{\@setfontsize\scriptsize{6}{7}}
\renewcommand\tiny{\@setfontsize\tiny{5}{6}}
\renewcommand\large{\@setfontsize\large{9.5}{11.5}}
\renewcommand\Large{\@setfontsize\Large{12}{14}}
\makeatother
\normalsize
\usepackage{xcolor}
\usepackage{tikz,pgfplots}
\usetikzlibrary{calc,arrows.meta,decorations.pathmorphing,bending}
\pgfplotsset{compat=1.18}
\IfFontExistsTF{Arimo-Regular.ttf}%
  {\newfontfamily\figsf{Arimo}[Extension=.ttf,UprightFont=*-Regular,BoldFont=*-Bold,ItalicFont=*-Italic,BoldItalicFont=*-BoldItalic]}%
  {\newfontfamily\figsf{texgyreheros}[Extension=.otf,UprightFont=*-regular,BoldFont=*-bold,ItalicFont=*-italic,BoldItalicFont=*-bolditalic]}
\makeatletter\begingroup\figsf\xdef\figsffam{\f@family}\endgroup\makeatother
\usepackage[subdued,italic,defaultmathsizes]{mathastext}
\MTDeclareVersion{figsans}{TU}{\figsffam}{m}{n}
\newcommand{\revon}{}\newcommand{\revoff}{}\newcommand{\chgon}{}\newcommand{\chgoff}{\color{black}}\newcommand{\mkon}{}\newcommand{\mkoff}{\color{black}}\newcommand{\rson}{}\newcommand{\rsoff}{\color{black}}\newcommand{\ron}{}\newcommand{\roff}{\color{black}}\newcommand{\pon}{}\newcommand{\poff}{\color{black}}\newcommand{\fon}{}\newcommand{\nmk}[1]{#1}\newcommand{\smk}[1]{#1}\newcommand{\umk}[1]{#1}\newcommand{\wmk}[1]{#1}\newcommand{\xmk}[1]{#1}\newcommand{\ymk}[1]{#1}\newcommand{\cmk}[1]{#1}\newcommand{\cmkcolor}{}\newcommand{\dmk}[1]{#1}\newcommand{\emk}[1]{#1}%
\newcommand{\hmk}[1]{#1}%
\newcommand{\imk}[1]{#1}%
\newcommand{\foff}{\color{black}}%
\definecolor{saRed}{RGB}{0,0,0}
\definecolor{saGreen}{RGB}{0,110,46}
\newcommand{\acmk}[1]{#1}%
\newcommand{\admk}[1]{#1}%
\newcommand{\aemk}[1]{#1}%
\newcommand{\afmk}[1]{#1}%
\newcommand{\agmk}[1]{#1}%
\newcommand{\ahmk}[1]{#1}%
\newcommand{\akmk}[1]{#1}%
\newcommand{\ajmk}[1]{#1}%
\newcommand{\aimk}[1]{#1}%
\newcommand{\almk}[1]{#1}%
\newcommand{\ammk}[1]{#1}%
\newcommand{\anmk}[1]{#1}%
\newcommand{\aomk}[1]{#1}%
\newcommand{\apmk}[1]{#1}%
\newcommand{\armk}[1]{#1}%
\newcommand{\aqmk}[1]{#1}%
\newcommand{\asmk}[1]{#1}%
\newcommand{\atmk}[1]{#1}%
\newcommand{\aumk}[1]{#1}%
\newcommand{\awmk}[1]{#1}%
\newcommand{\axmk}[1]{#1}%
\newcommand{\aymk}[1]{#1}%
\newcommand{\bbmk}[1]{#1}%
\newcommand{\bcmk}[1]{#1}%
\newcommand{\bamk}[1]{#1}%
\newcommand{\azmk}[1]{#1}\newcommand{\bdmk}[1]{#1}\newcommand{\bemk}[1]{#1}%
\newcommand{\kmk}[1]{{\textcolor{black}{#1}}}%
\newcommand{\lmk}[1]{{\textcolor{black}{#1}}}\newcommand{\gmk}[1]{{\textcolor{black}{#1}}}%
\newcommand{\jmk}[1]{{\textcolor{black}{#1}}}\newcommand{\qmk}[1]{#1}%
\usepackage{titlesec}
\titleformat{\section}
	{\normalfont\sffamily\bfseries\fontsize{9.5}{11.5}\selectfont\raggedright}
	{}{0pt}{}
\titleformat{\subsection}
	{\normalfont\sffamily\bfseries\fontsize{8.5}{10.5}\selectfont\raggedright}
	{}{0pt}{}
\titlespacing*{\section}{0pt}{10pt plus 2pt minus 1pt}{2pt}
\titlespacing*{\subsection}{0pt}{8pt plus 2pt minus 1pt}{1pt}

\usepackage[font={small,sf},labelfont={bf,sf},labelsep=period,justification=justified,singlelinecheck=false,skip=5pt]{caption}
\usepackage{fancyhdr}
\date{}

\makeatletter
\renewcommand{\fnum@figure}{\textbf{Fig. \thefigure}}
\renewcommand{\fnum@table}{\textbf{Table \thetable}}
\makeatother

\usepackage{scicite}

\makeatletter
\let\saOldThebibliography\thebibliography
\renewcommand{\thebibliography}[1]{\saOldThebibliography{#1}%
  \sffamily\footnotesize\raggedright
  \setlength{\itemsep}{0.5pt}\setlength{\parsep}{0pt}\setlength{\parskip}{0pt}}
\makeatother

\usepackage{url}
\def\scititle{
	Latent resonances in oscillator networks
}

\def\sciauthors{\aymk{Junjie Xiao$^{1\ast}$, Renaud Lambiotte$^{2}$, Hiroshi Kori$^{1}$}}%
\def\sciaffils{$^{1}$Department of Complexity Science and Engineering, Graduate School of Frontier Sciences, The University of Tokyo, Kashiwa, Chiba 277-8561, Japan. $^{2}$Mathematical Institute, University of Oxford, Oxford OX2 6GG, UK. \\ $^{\ast}$Corresponding author. Email: junjie-xiao@edu.k.u-tokyo.ac.jp}%


\begin{document}

\makeatletter
\twocolumn[
\begin{@twocolumnfalse}
\thispagestyle{fancy}
\noindent\begin{minipage}{\textwidth}
	\raggedright
	{\sffamily\fontsize{21}{24}\selectfont\bfseries\boldmath\color{saRed} \scititle\par}
	\vspace{0.8em}
	{\sffamily\fontsize{10.5}{13}\selectfont \sciauthors\par}
	\vspace{1.0em}
	{\sffamily\fontsize{9}{12}\selectfont%
	\rightskip=0pt\parfillskip=0pt plus 1fil\relax
\aymk{In networks of coupled oscillators, from neural circuits to power grids, damping is expected to suppress oscillations because of its dissipativity. Here we show that local damping can, counterintuitively, amplify oscillations along some \bdmk{input-output} channels, while the \bdmk{network's} global oscillation energy decreases and its spectral properties improve. The amplified oscillation arises from a latent resonance that lies in the spectrum yet barely manifests. The added passive damping reshapes this resonance by mixing it with adjacent eigenmodes. In random power-grid networks, damping can boost weakly responding channels by more than an order of magnitude, while subdominant channels can grow to exceed the originally strongest channel. Damping thus not only suppresses oscillations but also provides a way to redistribute oscillatory energy. For any passive network, the amplified response and its upper bound follow exactly from baseline quantities, offering a principled route for network design and control.}\par}%
\end{minipage}
\vspace{1.8em}
\end{@twocolumnfalse}
]
\makeatother

\section*{Introduction}
{\let\thefootnote\relax\footnotetext{\sffamily\fontsize{7}{8.5}\selectfont \sciaffils}}%

\noindent
\mkon{}Collective dynamics emerge in networks of coupled oscillators across nature and technology, from circadian cells and neurons\cite{Winfree1967,Acebron2005} to ecological populations\cite{Blasius1999}, mechanical structures\cite{Whitehead1998,CastanierPierre2006} and power grids\ymk{\cite{Filatrella2008}}. \xmk{Their function depends on how these networks withstand and respond to perturbations}\xmk{\cite{Hatziargyriou2021,Witthaut2022,Montanari2026}}. \aymk{Whether a network withstands small perturbations is commonly assessed from the eigenvalues of its linearized dynamics}\cite{Kundur1994}\aymk{. A stable network can still respond strongly. A periodic disturbance near the frequency of a mode drives a resonant response, as in the forced oscillations that can spread across an entire power grid}\cite{Follum2016,Maslennikov2017,NERC2019}\aymk{. Even when every mode decays fast, non-normal interactions can amplify perturbations}\cite{Trefethen1993,Asllani2018,Nazerian2024}\aymk{, as quantified by the reactivity and the pseudospectra of the system matrix}\cite{Neubert1997,TrefethenEmbree2005,Duan2022}\aymk{. How strongly a network responds also depends on where it is driven and where it is observed, the input-output channel}\cite{Nazerian2026}\aymk{, and on its topology}\cite{Tyloo2019,Zhang2019}.\mkoff{}

\aymk{In power grids, oscillations manifest as oscillating power flows on transmission lines, bounded by the thermal ratings and protection schemes of these lines. The grid dynamics follow the swing equation}\cite{Kundur1994}\aymk{ and can be described by a second-order Kuramoto model, in which these lines define the coupling between nodes}\cite{Filatrella2008}\aymk{. \bdmk{The physical lines} are fixed once built, but converter controls can introduce inertia and damping at selected nodes}\cite{DArcoSuul2014}\aymk{. For an isolated weakly damped oscillator, added damping shifts the mode away from the imaginary axis and reduces the forced response across all frequencies. Because damping dissipates energy, it is commonly assumed to suppress oscillations across the entire network. Inertia and damping have been allocated to minimize an $H_2$ performance measure of the grid}\cite{Poolla2017}\aymk{, to stabilize the synchronous state}\cite{Molnar2021}\aymk{, and, in our previous work, to suppress the worst-case oscillation}\cite{Xiao2026}\aymk{. Modal methods and screening indices guiding such placement rank modes, nodes, and channels according to their baseline behavior}\cite{Kundur1994,Pagnier2019,Tyloo2019}\aymk{, so weakly responding channels receive low priority. \bbmk{This practice treats added damping as a benefit, to be placed where it helps most.} The energy reasoning underlying this perspective, however, focuses on the total stored energy of the network, rather than its distribution over branches, which are the connections between nodes. Added damping can therefore improve global network behavior, as widely anticipated, without benefiting every channel, and may even amplify oscillations along some channels. It remains open how network structure and parameter allocation jointly determine which channels are amplified, and the magnitude of such amplification.}

\begin{figure}[!t]
\centering
\includegraphics{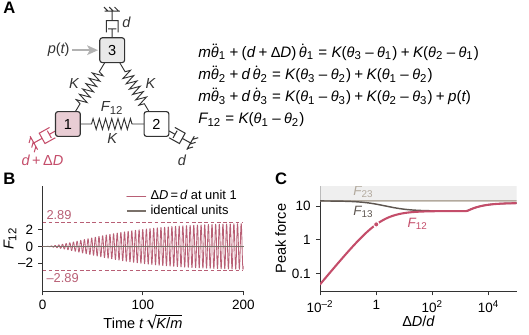}%
\caption{\textbf{\admk{A latent channel in three identical units and its opening by damping.}}
  (\textbf{A}) Three identical units on a triangle, drawn as masses coupled by springs, with the equations of motion. A disturbance $p(t)$ enters at unit 3, the \armk{force $F_{12}$ in the spring between units 1 and 2} is monitored, and the damping of unit 1 is raised by $\Delta D$.
  (\textbf{B}) Response of $F_{12}$ to a unit sinusoidal disturbance at \aumk{the resonance $\omega_0=\sqrt{3K/m}$ of the identical units}, started from rest. The red trace is for $\Delta D=d$ at unit 1, where the amplitude settles at \aumk{$2.89$, $20\%$ of the largest force with identical units}, and the gray trace for identical units, where $F_{12}$ stays zero. Dashed lines show the closed-form steady-state amplitude (section S2).
  (\textbf{C}) Largest steady \armk{force in each spring} over \aemk{the band of the oscillatory modes (Materials and Methods)}, for a unit sinusoidal disturbance at unit 3, as the damping of unit 1 is raised: $F_{12}$ (red), $F_{13}$ (dark gray) and $F_{23}$ (light gray). \aumk{$F_{23}$ stays at the largest force with identical units.} The gray band marks \armk{forces} larger than the largest \armk{force} with identical units, and the dot marks the case of (B). Parameters are \aumk{$m=1$, $K=1$ and $d=0.04$}, with $m$ and $d$ the inertia and damping of each unit.}%
  \label{fig:minimal}
\end{figure}

\begin{figure*}[!t]
  \centering
  \includegraphics[width=\textwidth]{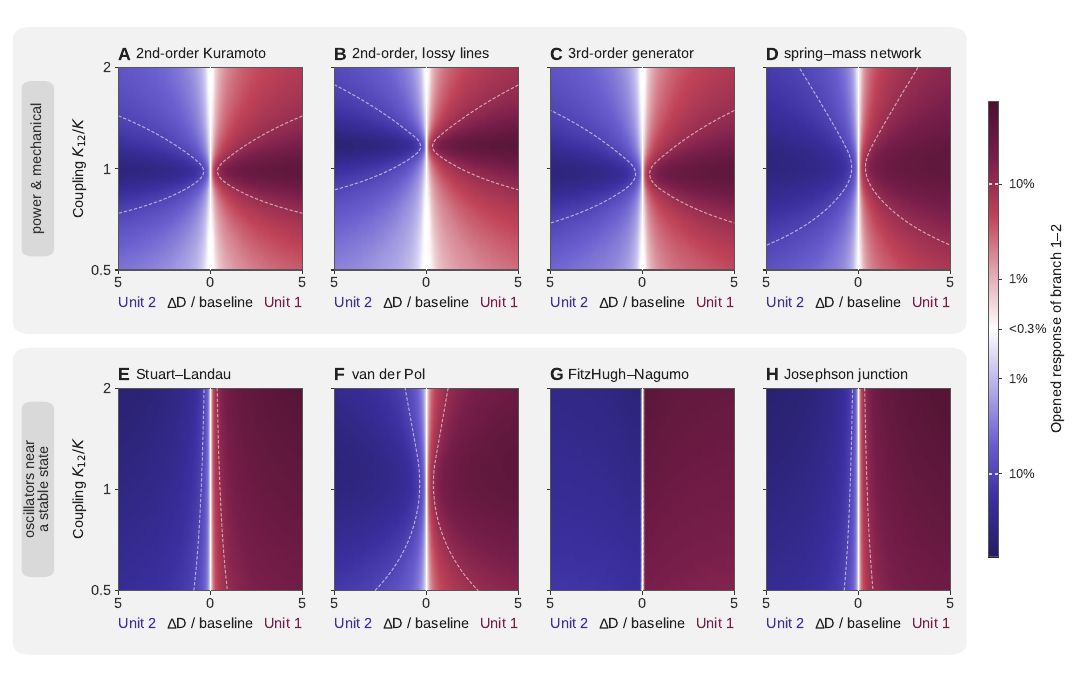}%
  \caption{\textbf{Opening of the latent resonance across oscillator models.}
  Three units on the triangle of Fig.~\ref{fig:minimal}A, with $K_{13}=K_{23}=K$, are driven at unit 3 and monitored through the flow on branch 1--2. Color shows the \akmk{opened response} of the channel, defined as the peak of the 1--2 flow over the band of the oscillatory modes (Materials and Methods) divided by the largest such peak among the three flows with identical units. The logarithmic color scale runs from $0.3\%$ (white) to $75\%$ and is common to all panels. The local dissipation of unit 1 (right half, rose) or unit 2 (left half, indigo) is raised by $\Delta D$, shown relative to its baseline value, and the coupling $K_{12}$ of the monitored branch is varied. The left half is the mirror image of the right, as the symmetry of units 1 and 2 requires \aymk{(section S2)}. The white line marks identical units ($\Delta D=0$), where the channel is latent and zero to within $10^{-9}$. The dashed line marks an \akmk{opened response} of $10\%$ (in G, within the width of the white line).
  (\textbf{A} to \textbf{D}) Power-grid and mechanical networks, with $\Delta D$ on the damping coefficient: second-order Kuramoto (A), whose \aumk{middle row, $K_{12}=K$,} is the network of Fig.~\ref{fig:minimal} with a steady power injection at unit 3, second-order Kuramoto with lossy lines (B), third-order generator (C) and spring--mass network (D).
  (\textbf{E} to \textbf{H}) Nonlinear oscillators linearized about a stable state: Stuart--Landau below the Hopf point, with $\Delta D$ on the decay rate (E); van der Pol below the onset of self-oscillation, on $|\mu_i|$ (F); FitzHugh--Nagumo near the Hopf point, on the recovery decay $\lambda_i$, which also shifts the rest state (G); and underdamped Josephson junctions, on the damping (H). Contours are from cubic interpolation of the computed grid. Equations and parameters are given in table~\ref{tab:osc_models}.}
  \label{fig:universality}
\end{figure*}

\aymk{Reorganizing this oscillatory energy can open a channel \bdmk{on which a resonance of the network barely shows up at baseline}. We call such a resonance \emph{latent} on that channel, and the channel a latent channel. Figure~\ref{fig:minimal} shows the simplest case. The units are coupled by springs, and we monitor the spring forces acting on the inter-unit connections. For three identical units arranged in a triangle, a disturbance at unit 3 excites a lightly damped resonance, yet the force between units 1 and 2 vanishes at all frequencies, because the two units are interchangeable}\bbmk{\cite{RubinMeadows1972,Whalen2015}}\aymk{. Increasing the damping of unit 1 brings this resonance into the force signal, even as the total network damping rises. We say that the damping opens the channel. Any mismatch between units 1 and 2, in damping, inertia, or coupling, opens the channel, whereas identical parameter changes do not. More generally, \bbmk{identically positioned nodes with the same inertia and damping move together when the disturbance enters elsewhere, so latent channels of this kind are set by the structure of the network}}\bbmk{\cite{OClery2013,Schaub2016}}\aymk{. \bbmk{Unlike resonant and non-normal amplification, which are properties of a given network, this amplification is caused by a change of the network itself.}}

\aymk{\bbmk{Here we derive exactly the response of any channel to a local change of damping from the frequency response of the network before the change.} For any passive network, both the resulting channel response and its upper bound can be predicted prior to modifying damping. \aymk{A modal expansion} reveals that inter-node damping heterogeneity mixes adjacent eigenmodes, increasing the spectral weight of a latent resonance on the corresponding channel. In random power-grid networks, local damping can amplify oscillations on weakly responding channels \azmk{by more than an order of magnitude}, while subdominant branches may develop responses exceeding the \bdmk{network's} largest baseline oscillation. This latent-channel mechanism also operates in \bbmk{eight oscillator models} (Fig.~\ref{fig:universality}), the pan-European transmission grid \azmk{(section S6)}, and under local reductions of node inertia or line coupling (section \bcmk{S12}). Our theory yields an important implication for collective dynamics in coupled oscillator networks: a more stable network is not necessarily a safer one. Stability is governed by the spectrum, whereas safety depends on how network structure and parameters jointly distribute the response across its connections.}

\section*{Results}

\subsection*{Model and response channels}

\admk{We consider a linear network driven at one input and observed at one output, whose state $x$ evolves according to}
\begin{equation}
\dot x = A\,x + \hmk{\hat b}\,p(t) ,\qquad y = \hmk{\hat c}^{\top}x \admk{,}
\label{eq:statespace}
\end{equation}
\wmk{\admk{where the disturbance $p(t)$ enters through $\hat b$ and the output $y$ is read through $\hat c$. Its} response to a sinusoidal drive of frequency $\omega$ is the scalar \cmk{transfer function} $\hmk{\hat c}^{\top}(j\omega I-A)^{-1}\hmk{\hat b}$\admk{.}}

\xmk{For networks of \emk{$N$} masses coupled by springs, as in Fig.~\ref{fig:minimal}A, \admk{Newton's second law gives $M\ddot\theta+D\dot\theta+L\theta=b_r\,p(t)$ for the displacements $\theta$. Here $M$ and $D$ are the diagonal mass and damping matrices, and $L$ is the symmetric stiffness matrix, a weighted network Laplacian. The vector $b_r=e_r$ selects the driven node $r$. These equations map to Eq.~\ref{eq:statespace} with $x=(\theta,\dot\theta)$, $\hat b=(0,M^{-1}b_r)$ and $\hat c=(c_\ell,0)$, where $c_\ell$ reads out the flow on a monitored branch $\ell${, the force in a spring or the power on a line}. For a branch between nodes $i$ and $j$, $c_\ell=-L_{ij}(e_i-e_j)$.} \cmk{The channel from node $r$ to branch $\ell$ is then}}
\begin{equation}
G_{r,\ell}(j\omega)
= c_\ell^{\top}\,Z(j\omega)^{-1}\,b_r
= \sum_{k=1}^{2N}\frac{R_k}{\,j\omega-\cmk{s_k}\,} ,
\label{eq:channel}
\end{equation}
\xmk{where} $Z(j\omega)=-\omega^{2}M+j\omega D+L$ \xmk{is the dynamic stiffness matrix}\admk{.}

\admk{The poles $s_k$ of Eq.~\ref{eq:channel} are the eigenvalues of the network and are the same for every channel. The residues $R_k$ depend on the channel. The poles and the mode vectors $v_k$ solve the quadratic eigenvalue problem $(s_k^2 M+s_k D+L)\,v_k=0$. Because $M$, $D$ and $L$ are symmetric, the left mode vectors equal the right ones, $w_k=v_k$.} \xmk{The residues are} $R_k=\bigl(c_\ell^{\top}v_k\bigr)\bigl(w_k^{\top}b_r\bigr)\big/\bigl(v_k^{\top}(2s_k M+D)\,v_k\bigr)$\admk{. The output factor $c_\ell^{\top}v_k$ measures how much of mode $k$ reaches the output, and the input factor $w_k^{\top}b_r$ measures how strongly the input excites it}. %
\admk{A lightly damped pole is a resonance of the network. A channel with a large residue at this pole responds strongly near its frequency.} \xmk{\admk{A channel with a small residue barely shows the same resonance}\lmk{\cite{Kalman1960,Rosenbrock1970}}. \admk{\ajmk{A change} of parameters can raise this residue, and the channel then opens}.} \admk{In the canonical case of uniform inertia and damping, $M=mI$ and $D=dI$, the dynamics reduce to the damped wave equation on the graph}\cite{Caputo2013}\admk{. The mode vectors are then the eigenvectors of $L$ and do not depend on $d$. Uniform damping therefore moves the poles but leaves the input and output factors unchanged, and a latent channel stays latent. Opening it requires damping that differs between nodes.}

\subsection*{\cmk{Opening of a latent channel by passive damping}}

\admk{Figure~\ref{fig:minimal} illustrates that, in three identical units on a triangle, the channel from unit 3 to branch 1--2 is exactly zero.} \admk{Units 1 and 2 are coupled to unit 3 in the same way, so {the modes can be chosen such that} each either moves them together (an in-phase mode) or in opposite directions (an antiphase mode). An in-phase mode has equal components at units 1 and 2, so its output factor on branch 1--2 is zero. An antiphase mode has no component at unit 3, so the drive cannot excite it. Every residue of the channel therefore vanishes, and the channel is zero at every frequency}\cite{RubinMeadows1972,Whalen2015}.

{{{The channel follows the difference between the dampings of units 1 and 2.} Damping added at unit 1 creates this difference. {The resonant motion} then no longer moves the two units exactly together, and branch 1--2 carries {a force} at {this} resonance} (Fig.~\ref{fig:minimal}B). {Let $Q$ exchange units 1 and 2. Because \armk{$Q^\top MQ=M$, }$Q^\top LQ=L$, $Qb_r=b_r$ and $Q^\top c_\ell=-c_\ell$, the exchange maps the network with dampings $(D_1,D_2)$ onto the one with $(D_2,D_1)$. It leaves the eigenvalues unchanged and reverses the sign of the channel. At a given mean damping, \aumk{the channel is therefore an odd function of $D_1-D_2$}. To leading order,}}%
\begin{equation}
  \begin{aligned}
  G_{r,\ell}(s)&=\tfrac12\,h(s)\,(D_1-D_2)+O\big((D_1-D_2)^{3}\big).
  \end{aligned}
  \label{eq:mirror}
\end{equation}
\cmk{\admk{Here \aumk{$h$ refers} to the balanced network, in which $D_1$ and $D_2$ are both set to their mean.\aumk{ In this network the drive moves units 1 and 2 equally. Making their dampings differ adds damping forces of equal size and opposite sign at the two units, in proportion to $D_1-D_2$. Branch 1--2 reads the difference of their motions, so the effects of these opposite forces add on this branch, and the channel grows in proportion to $D_1-D_2$. The factor $h$ is the product of how strongly the drive moves unit 1 and how strongly a force at unit 1 reaches the output}} (section S2). \aumk{Damping added at unit 1 leaves the spectral margin\aymk{, the largest real part among the nonzero eigenvalues,} unchanged, because one of the two resonances at $\omega_0=\sqrt{3K/m}$ does not move unit 1} (section S2)\admk{.}\lmk{}}

\lmk{\admk{The symmetry of the three units is a special case of a more general structure. A grouping of the nodes into cells is an \ahmk{externally} equitable partition if every node of a cell has the same total coupling to each other cell}\cite{Rahmani2009,OClery2013,Schaub2016}\admk{. Suppose that the disturbed node forms a cell of its own and that inertia and damping are equal within each cell. The disturbance then moves all nodes of a cell together, and no branch inside a cell carries flow} (Materials and Methods). The latent channels \awmk{that cells create} thus follow from \awmk{the} \admk{structure}\awmk{ of the network}, before any modal analysis\ajmk{ (fig.~\ref{fig:regular})}. Damping added at one node of a cell \admk{generically} opens the \admk{channels inside that cell. Damping added at a node that forms a cell of its own, or equally within every cell, keeps these channels at zero. When} the nodes of a cell differ slightly, as in the random power-grid networks below, \admk{these channels are weak rather than zero, and damping opens them} in the same way.}%

\begin{figure*}[!t]
  \centering
  \includegraphics[width=\textwidth,trim=0bp 8.6bp 0bp 10.9bp,clip]{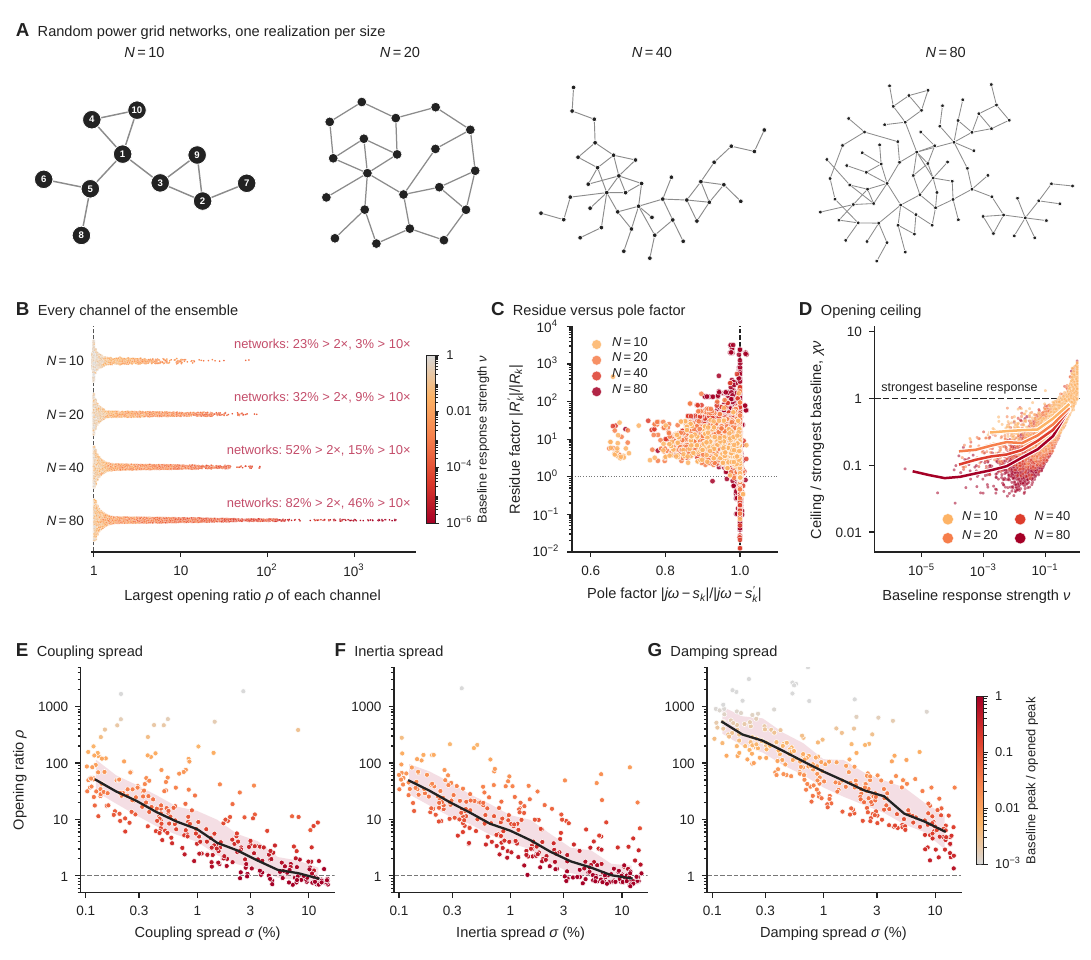}%
  \caption{\textbf{Latent channels in random power-grid networks and the conditions for \admk{large} opening.}
  \acmk{(\textbf{A}) One of the $300$ realizations at each size $N$ from the random growth model of ref.~\cite{Schultz2014}. Inertia, damping and coupling vary independently and uniformly within $\pm5\%$ of their means. Numbers label the nodes of the $N=10$ network\azmk{, whose parameters are listed in table~\ref{tab:ten_machine_parameters}}.
  (\textbf{B}) Largest opening ratio $\rho$ of every \afmk{disturbance--branch} channel ($3.60\times10^{6}$ in total) at the best damping node, with one row per size. Color shows $\nu$, relative to the strongest branch under the same disturbance, and vertical spread shows the density of channels. \afmk{Text at the right of each row gives} the share of networks with a channel opened by more than a factor of $2$ or $10$ when the damping of one node is raised by the network's mean damping. The number of channels per network grows with $N$, \ajmk{but} the share of channels opened by more than a factor of $2$ stays between $0.66\%$ and $1.01\%$ at every size. %
  (\textbf{C}) Residue factor $|R_k'|/|R_k|$ versus pole factor $|j\omega-s_k|/|j\omega-s_k'|$ at the resonance nearest the post-intervention worst-case frequency, for the $26{,}215$ channels opened by more than a factor of $2$. Primes denote values after the intervention, and the product of the two factors is the modal gain ratio. The dashed line marks an unchanged pole distance, and the dotted line an unchanged residue. Color shows $N$.
  (\textbf{D}) \aumk{Opening ceiling $\chi$ at the best damping node, multiplied by $\nu$ to express it relative to the strongest baseline response, versus $\nu$. The ceiling is read over $0.2$--$10$~Hz\azmk{, and dots show} $6000$ channels drawn at random per size.} Lines show the median of all channels of each size per half decade of $\nu$. The dashed line marks $\chi\nu=1$, at which the ceiling matches the strongest baseline response. Color shows $N$.
  (\textbf{E} to \textbf{G}) $\rho$ of the channel from node \armk{5} to \armk{$\ell_{4,10}$}, which joins two \admk{identically positioned} nodes of the $N=10$ network in (A), versus the log-normal spread $\sigma$ of the couplings (E), the \ahmk{inertias} (F) or the dampings (G), with the other two parameters identical. \afmk{The damping of node \armk{4} is raised by the mean damping.} Color shows the baseline peak relative to the opened peak at identical parameters. The line and band show the binned median and interquartile range, and the dashed line marks $\rho=1$.}}%
  \label{fig:ensemble}
\end{figure*}

\begin{figure*}[!t]
\centering
\includegraphics[width=192.1bp]{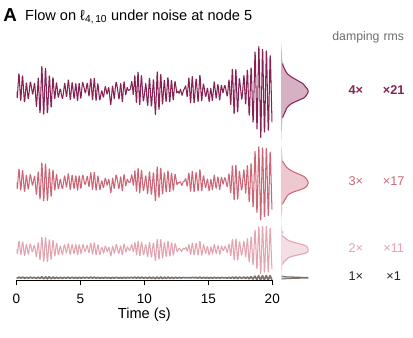}%
\includegraphics[width=150.2bp]{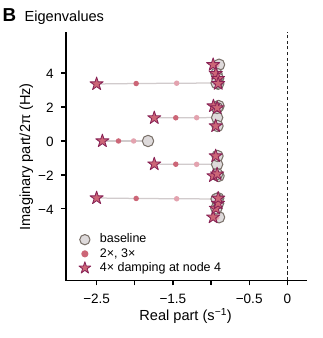}%
\includegraphics[width=160.6bp]{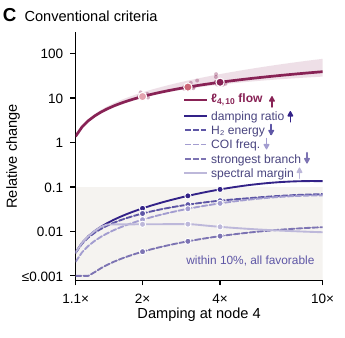}%
\\[2pt]%
\includegraphics[width=502.9bp]{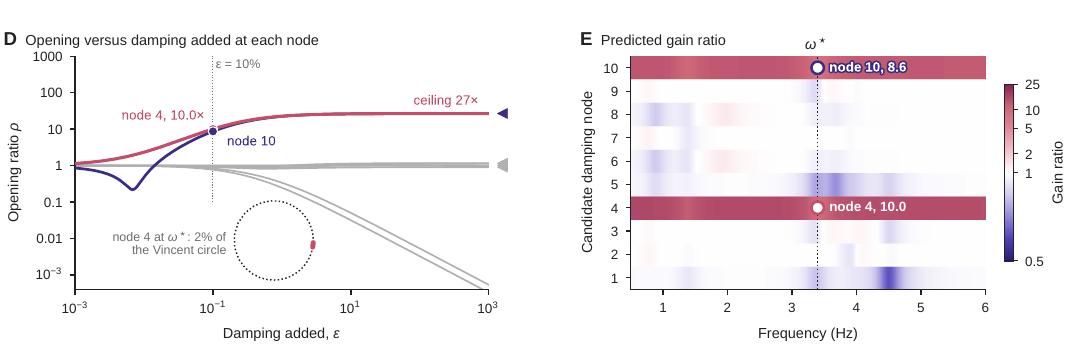}%
\caption{\textbf{Opening of a latent channel by passive damping at one node of a random ten-node network.}
  \acmk{Network of Fig.~\ref{fig:ensemble}A. The disturbance enters at node~\armk{5}, as band-limited white noise over $0.2$--$10$~Hz in \ammk{(A)} and as a unit sinusoid in \ammk{(D) and (E)}. Damping is added at node~\armk{4}, and the flow on branch \armk{$\ell_{4,10}$} is monitored.
  (\textbf{A}) Flow on \armk{$\ell_{4,10}$} with the damping of node~\armk{4} raised to $1$ to $4$ times its value, for one noise realization, relative to the baseline maximum. Traces are offset vertically on a common scale. The right panel shows the distribution of each trace over $30$~s, of which the first $20$~s are shown, and the rms relative to baseline for white noise over the band.
  (\textbf{B}) Eigenvalues at baseline (gray circles) and with $2$ and $3$ times (dots) and $4$ times (stars) the damping of node~\armk{4}. The dashed line marks the imaginary axis. \bdmk{At baseline, the} pair nearest the axis, at \armk{$4.50$}~Hz, sets the spectral margin.
  (\textbf{C}) \aumk{The $\ell_{4,10}$ flow (rose) compared with conventional criteria (indigo) as the damping of node~4 is raised: smallest damping ratio, spectral margin, energy stored under white noise at the disturbed node ($H_2$, section S4.3), rms frequency at the center of inertia, and rms flow on the branch strongest at baseline. Curves show the size of the relative change, $|X'-X|/X$, so that $10$ means $11$ times the baseline\aymk{.} \aymk{D}ashed lines mark a decrease, the gray area marks changes below $10\%$, and changes below $0.1\%$ are drawn at $0.1\%$. Criteria that depend on the disturbance are medians over the eight disturbance nodes other than 4 and 10, and the rose band shows the range of the $\ell_{4,10}$ flow over them.}
  (\textbf{D}) Opening ratio $\rho$ of the \armk{$\ell_{4,10}$} channel versus the damping added at each node as a fraction of the network total (red, node~\armk{4}; blue, node~\armk{10}; gray, other nodes). %
Triangles mark the closed-form ceilings as $\Delta D_i\to\infty$ (Eq.~\ref{eq:ceiling}), and dots mark $\varepsilon=10\%$. Inset: the gain at the baseline $\omega^\star$ for damping at node~\armk{4} in the complex plane, a Vincent circle\cite{GhandchiTehrani2006} (dotted), of which the passive range (red) covers about \armk{$2\%$}.
  (\textbf{\ammk{E}}) Gain ratio $|G_{r,\ell}(\Delta D_i;j\omega)|/|G^{0}_{r,\ell}(j\omega)|$ from Eq.~\ref{eq:mobius_closed_form} over damping node and frequency \afmk{for $\varepsilon=10\%$ at each node}, on a logarithmic color scale centered on $1$. The dotted line marks the baseline $\omega^\star$.}}
  \label{fig:local-damping-fragility}
\end{figure*}

\subsection*{\admk{Latent resonances across oscillator models}}

\rson{}\cmk{\admk{Latent resonances arise in the same way in other networks with lightly damped oscillatory modes}, because the \admk{symmetry} argument above does not depend on the dynamics of the units. We tested eight such models: four power-grid and mechanical networks}\cite{Filatrella2008,Schmietendorf2014,Montanari2026} \cmk{and four nonlinear oscillators linearized about a stable state}\cite{Pikovsky2001,FitzHugh1961,Montanari2026} \cmk{(Fig.~\ref{fig:universality}; equations in table~\ref{tab:osc_models}). \admk{In} every model the channel is zero with identical units. Dissipation added at unit 1 or unit 2 opens it by the same amount, and the opened response peaks at \admk{the frequency of} an oscillatory mode of the network. \jmk{The \akmk{opened response} is large} where the in-phase \emk{and antiphase modes} lie close in frequency}\cite{Dobson2001}\cmk{, because damping mixes two modes more strongly the closer they are. In the power-grid and mechanical models it peaks where the two modes cross, near $K_{12}=K$. When the coupling is weak compared with the restoring force of each unit, the two modes stay close for every $K_{12}$. The channel then opens over a wide range of $K_{12}$ for masses held by stiff ground springs, and at every $K_{12}$ for weakly or dissipatively coupled oscillators \kmk{(Fig.~\ref{fig:universality}, D to H)}.}\rsoff{}%

\subsection*{\wmk{Latent resonances in power grids}}

\aemk{With each synchronous generator described by its swing equation, a power grid becomes a network of second-order Kuramoto oscillators}\ymk{\cite{Kundur1994,Filatrella2008,NishikawaMotter2015}},%
\begin{equation}
M_i\ddot{\theta}_i+D_i\dot{\theta}_i
=\Omega_i+\hmk{\sum_n K_{in}\,g(\theta_n-\theta_i)}+p_i(t) \aemk{.}
\label{eq:network_dynamics}
\end{equation}
\admk{\aemk{Here} $\theta_i$ is the phase of node $i$ in the synchronously rotating frame, $M_i$ and $D_i$ are its inertia and damping, and $p_i(t)$ is a disturbance at node $i$. The coupling function $g$ is odd, with $g=\sin$ for an AC power grid. The constant injections $\Omega_i$ and the couplings $K_{in}=K_{ni}\ge 0$ together fix \aemk{a stable} operating point $\theta^{\mathrm{op}}$. \aemk{Linearized about $\theta^{\mathrm{op}}$, Eq.~\ref{eq:network_dynamics} has the stiffness matrix with the entries} $L_{in}=-K_{in}\,\mathrm{d}g/\mathrm{d}\theta$, evaluated at $\theta=\theta^{\mathrm{op}}_n-\theta^{\mathrm{op}}_i$, for $i\neq n$. The stiffness matrix thus depends on the operating point as well as on the topology. The output $c_\ell$ reads the linearized power flow on branch $\ell$.}
\admk{Each channel has a worst-case frequency $\omega^\star$, at which its gain is largest within the band of interest \aemk{$\mathcal{W}$ (Materials and Methods)}. \aumk{To measure how far an intervention opens a latent channel, we define the opening ratio of a channel as its peak gain after the intervention divided by that before,}
\begin{equation}
\rho=\frac{\Gamma}{\Gamma_0},\qquad \Gamma=\max_{\omega\in\mathcal{W}}\bigl|G_{r,\ell}(j\omega)\bigr|,
\label{eq:opening_ratio}
\end{equation}
\atmk{where $\Gamma_0$ is the same maximum at baseline, so that each peak is taken at its own worst-case frequency.} The baseline response strength $\nu$ is the channel's baseline peak divided by the largest baseline peak \aemk{over all} monitored branches under the same disturbance.}

\aemk{Passive damping opens latent channels in random power-grid networks. We \aemk{generate} networks of $10$ to $80$ nodes with the random growth model of Schultz, Heitzig and Kurths\cite{Schultz2014}, which reproduces the topology statistics of real power grids. Inertia, damping and coupling vary from node to node and from line to line by up to $5\%$ around their mean values (Fig.~\ref{fig:ensemble}A)\bamk{, as for a grid of nearly identical units such as inverter-based generators of one type and setting}. The damping of a single node is raised by the network's mean damping\aumk{, which about doubles the damping of that node}. In $82\%$ of the networks \aumk{of $80$ nodes}, this increment at some node opens some channel to more than twice its baseline peak. In $46\%$ it opens one by more than an order of magnitude (Fig.~\ref{fig:ensemble}B)\apmk{.} \azmk{Over all sizes, a channel opened to more than twice its baseline carries after the opening up to $27\%$ of the strongest baseline flow under the same disturbance, and one opened by more than an order of magnitude up to $21\%$.} \aemk{Openings by more than an order of magnitude} lie between two nearly interchangeable nodes. \afmk{\aymk{All} of them are on a branch inside a cell of the \ahmk{externally} equitable partition, most on a branch whose two ends share all their other neighbors, with the damping at one end of that branch (table~\ref{tab:twin_control}).} At $N=80$, $65\%$ of the networks \apmk{that contain such a pair of nodes} hold a channel opened by more than an order of magnitude, against $6\%$ of the others (table~\ref{tab:twin_control}). \aumk{The increment can also raise the strongest flow itself markedly, by up to $14\%$. It also amplifies $2913$ other channels beyond the strongest baseline flow\aymk{.} \aymk{A}ll of them start above $85\%$ of it and grow by at most $22\%$. Four times this increment amplifies the channels opened by more than an order of magnitude to up to $39\%$ of the strongest baseline flow, and channels starting from as low as $69\%$ of the strongest baseline flow then exceed this flow.}}%

\admk{\aemk{The ten-node network of Fig.~\ref{fig:ensemble}A shows how one increment raises some channels and lowers others. The disturbance enters at node \armk{5}, and} the flows on all \armk{$11$} branches are monitored (Fig.~\ref{fig:local-damping-fragility}). Nodes \armk{4 and 10} are linked to each other and to node \armk{1} only, so the two are positioned identically in the network. \aemk{They form a nearly interchangeable pair of the kind found in the ensembles.} As in the ensembles, the network's mean damping ($\varepsilon=10\%$ of its total damping) is added at one node at a time, with inertia and all other parameters held fixed.}
\admk{\aemk{\apmk{The same} increment at node \armk{5} lowers the \armk{strongest channel, $\ell_{3,9}$, by $25\%$}.} The \armk{$\ell_{4,10}$} channel is the weakest of the \armk{eleven} at baseline. Adding the increment at node \armk{4} raises it \aemk{to \armk{$10.0$} times its baseline} (Fig.~\ref{fig:local-damping-fragility}\ammk{D}). \apmk{None} of the other \armk{ten} rises by more than \armk{$7\%$}. Doubling the damping of node \armk{4} is nearly the same increment. Under band-limited white noise at node \armk{5}, it raises the rms flow on \armk{$\ell_{4,10}$} \aemk{to \armk{$11$} times its baseline} (Fig.~\ref{fig:local-damping-fragility}A). \aemk{The spectral margin \armk{improves by $1.4\%$}} (Fig.~\ref{fig:local-damping-fragility}\ajmk{, B and C}). \aemk{As the damping of node~\armk{4} is raised up to four times its value, the other global indicators change by less than $10\%$} (Fig.~\ref{fig:local-damping-fragility}C).}

\admk{The largest openings fall on the weakest channels. \aemk{With the disturbance placed at each of the ten nodes in turn, the network has \armk{$110$} disturbance--branch channels. The eight weakest at baseline} are the flows on \armk{$\ell_{4,10}$} with the disturbance at a node other than \armk{4 and 10}. \aemk{These eight also open the most, with damping at node \armk{4}. When the disturbance enters at node \armk{4 or 10},} the same branch carries the strongest flow of the network. \aemk{Whether a resonance appears on a channel thus depends on both its input and its output.} \armk{No other channel opens by more than a factor of $1.3$} (fig.~\ref{fig:hidden-response-channels}C). For the disturbance at node \armk{5}, \armk{three of four} conventional screening indices place the \armk{$\ell_{4,10}$} channel last among the \armk{eleven, and the fourth places it seventh} (fig.~\ref{fig:competing-indices}). These indices are the baseline peak gain, the in-band $H_2$ energy\cite{Tyloo2019}, the residue of the least-damped in-band mode\cite{Kundur1994} and the electrical distance from the disturbance.}

\aemk{The pan-European transmission grid is far more heterogeneous than the random networks. Only the $468$ of its $3809$ buses that host a producing generator have inertia. \apmk{The inertia of these buses, the nodal damping and the line coupling each vary by more than an order of magnitude} (section S6). Almost no two buses are identically positioned. Raising the damping of a single generator bus to three times its value, with inertia and dispatch unchanged, \azmk{can make a line the strongest of the grid. Under a disturbance at G\"{o}sgen (Switzerland), tripling the damping at K\"{u}hmoos (Germany) raises the G\"{o}sgen--Flumenthal line from $89\%$ to $104\%$ of the strongest baseline flow and lowers the strongest line to $54\%$ (fig.~\ref{fig:real-grid}H). \bdmk{Tripling the damping of a single bus} also opens latent channels, the largest by a factor of \bdmk{about $3$} (fig.~\ref{fig:real-grid}D).} \atmk{For a typical generator bus, the increment is about $0.15\%$ of the total damping of the grid} (section S6). The spectral margin changes by less than $1\%$.}%

\revon{}

\subsection*{\xmk{Origin and prediction of the opening}}

\admk{\afmk{In the ten-node network, damping at node \armk{4} opens the channel at the \armk{$3.37$}~Hz resonance. At baseline, a disturbance at node \armk{5} drives this resonance strongly. The resonance, however, moves nodes \armk{4 and 10} together with nearly equal amplitudes, so the two ends of \armk{$\ell_{4,10}$} hardly move relative to each other and the branch carries little flow. The \armk{$3.42$}~Hz resonance moves the two nodes in opposite directions and would carry flow on the branch, but it has almost no component at node \armk{5}, so the disturbance barely drives it. The channel is therefore weak at baseline, and because the two nodes differ slightly, it is not zero.}} %

\admk{Near resonance the modal term peaks at about $|R_k|/|\mathrm{Re}\,s_k|$. A latent resonance has a small peak at baseline\afmk{, and any} change that raises $|R_k|$ by a larger factor than $|\mathrm{Re}\,s_k|$ raises \afmk{it}. Adding damping leaves $M$, $L$ and the ports unchanged, so it can change the numerator of the residue only by moving the mode vectors. Under proportional or, more generally, Caughey damping\cite{CaugheyOKelly1965}, the mode vectors are fixed by $M$ and $L$ alone. A damping increment at a single node generically \aemk{breaks this condition and} moves the mode vectors\cite{Montanari2026}, so a passive increase of damping can open a latent channel. \armk{To first order, each} other mode mixes into the resonance with a weight inversely proportional to its spacing from the resonance\aemk{ (eq.~\ref{eq:sup_eigvec_perturbation})}. \ajmk{The first-order shift of the pole}, in contrast, \aemk{depends only on the component of the resonance at the damped node} \aemk{(section S4.1)}. \afmk{In the ten-node network,} damping at node \armk{4} mixes the \armk{$3.42$}~Hz \afmk{mode}\afmk{, which carries by far the most flow on \armk{$\ell_{4,10}$} (section S4.1),} into the \armk{$3.37$}~Hz resonance\afmk{.} This mixing raises the output factor that the near symmetry of the pair had suppressed and leaves the input factor almost unchanged.}%

\admk{With the network's mean damping added at node \armk{4}, the residue of the \armk{$3.37$}~Hz resonance on \armk{$\ell_{4,10}$} grows by a factor of \armk{$7.7$}. \armk{The $3.42$~Hz resonance lies \bdmk{$0.044$}~Hz above it, and its residue on this channel has nearly the same size and the opposite sign, so at baseline the two terms nearly cancel. Its residue grows by the same factor, and the} added damping also moves \armk{its} pole away from the driving frequency (Fig.~\ref{fig:local-damping-fragility}B)\afmk{\armk{, which weakens the cancellation, so}} the channel opens by a factor of \armk{$10.0$}. \aemk{\afmk{The ensembles} follow the same pattern. In \afmk{$92\%$} of the channels opened to more than twice their baseline, the residue of the resonance at the opened peak rises \afmk{while} its pole does not come closer to the driving frequency (Fig.~\ref{fig:ensemble}C). At baseline, \atmk{as on \armk{$\ell_{4,10}$}, }the output factor of this resonance \apmk{is typically far smaller than its input factor} (section S4.2).}}%

\admk{Adding damping $\Delta D_i>0$ at a single node $i$ \afmk{inserts a damper that only absorbs energy (Materials and Methods) and} changes the dynamic stiffness matrix by the rank-one term $j\omega\,\Delta D_i\,e_ie_i^\top$\aymk{.} The Sherman--Morrison formula\cite{ShermanMorrison1950} then gives the channel gain in closed form as a function of the damping increment,}
\begin{equation}
  G_{r,\ell}(\Delta D_i;j\omega)
  =G^{0}_{r,\ell}-\frac{j\omega\,\Delta D_i\,\beta_i\gamma_i}{1+j\omega\,\Delta D_i\,\alpha_i} .
\label{eq:mobius_closed_form}
\end{equation}
\admk{Every coefficient comes from a single evaluation of the baseline receptance matrix $Z_0^{-1}(j\omega)$, \afmk{the inverse of the dynamic stiffness matrix}\cite{MottersheadRam2006}. The baseline channel gain $G^{0}_{r,\ell}=c_\ell^\top Z_0^{-1}b_r$ is small on a latent channel. \afmk{A unit drive at $r$ moves node $i$ by $\gamma_i=e_i^\top Z_0^{-1}b_r$, and the added damper turns this motion into a force at node $i$. This force reaches output $\ell$ through $\beta_i=c_\ell^\top Z_0^{-1}e_i$ and acts back on node $i$ through its driving-point receptance $\alpha_i=e_i^\top Z_0^{-1}e_i$, which gives the denominator.} The formula is exact. \aymk{\azmk{With symmetric $M$, $D$ and $L$ and positive damping, the driving-point admittance $j\omega\alpha_i$ of the damped node is positive real, and we call the network passive.} P}assivity keeps its denominator from vanishing at any damping (Materials and Methods). As $\Delta D_i$ grows from $0$ to $\infty$, $G_{r,\ell}$ \afmk{moves along} an arc of a M\"obius circle in the complex plane\afmk{, and passivity fixes the reachable part of this arc and thus bounds the opening. In structural dynamics this circle is known as the Vincent circle of a rank-one modification, where it is used to \aqmk{design modifications that }place zeros}\aqmk{ or minimize a response}\aqmk{\cite{MottersheadRam2006,GhandchiTehrani2006}}.\atmk{ \aymk{The} same rank-one form also applies when the inertia of node $i$ is changed, with $j\omega\Delta D_i$ replaced by $-\omega^2\Delta M_i$ (section S12), and when the coupling of a single branch is changed, with $j\omega\Delta D_i$ replaced by the change of its coupling and $e_i$ by the difference of the unit vectors of its two end nodes.}}\aymk{ Equation~\ref{eq:mobius_closed_form} is linear in $b_r$, so it also holds with $b_r$ replaced by any input vector and, by superposition, for disturbances at several nodes. When the damping or inertia of several nodes, or the coupling of several branches, is changed at once, the scalar denominator becomes a small matrix, and the gain still follows from $Z_0^{-1}$ alone (Materials and Methods).}%

\admk{Equation~\ref{eq:mobius_closed_form} thus predicts the opening from baseline quantities only. For each candidate node $i$ and a given increment $\Delta D_i$, the peak of $|G_{r,\ell}(\Delta D_i;j\omega)|$ over \aemk{$\mathcal{W}$}, and with it $\rho$, follows from the same $Z_0^{-1}$ without rebuilding $Z$. The opened peak is set by $\beta_i\gamma_i/\alpha_i$ and changes little with the small asymmetry of the network, whereas $\rho$ is inversely proportional to the baseline peak, which that asymmetry sets.}%

\admk{With the same amount of damping, node \armk{4 opens the channel by a factor of $10.0$ and node 10 by $8.6$}, whereas any other node changes it by at most \armk{$23\%$} (Fig.~\ref{fig:local-damping-fragility}D). \afmk{The modal view ranks the nodes the other way.} Damping at node \armk{5 or node 1} raises the damping ratio of the \armk{$3.37$}~Hz resonance by \aymk{$24\%$ and $20\%$}\afmk{, and damping at node \armk{4 or node 10} by \atmk{less than $2\%$}} \ammk{(section S5)}.}%
\admk{\afmk{In Eq.~\ref{eq:mobius_closed_form}, at} the baseline $\omega^\star$, damping at node \armk{4} and damping at node \armk{10} add corrections of nearly equal magnitude and opposite sign\afmk{, because exchanging} the two nodes reverses $\beta_i$ and leaves $\gamma_i$ and $\alpha_i$ nearly unchanged. Both corrections are much larger than the baseline gain, so both nodes open the channel. The pointwise gain ratio \afmk{over} every damping location and frequency shows that nodes \armk{4 and 10} open the channel and that the other nodes leave it nearly unchanged (Fig.~\ref{fig:local-damping-fragility}\ammk{E}).}

\subsection*{\xmk{Ceiling of the opening and parameter heterogeneity}}
\admk{\afmk{Equation~\ref{eq:mobius_closed_form} also} bounds the opening. As the damping increment grows without limit, the factor $j\omega$ cancels, and \afmk{the gain} converges to}
\begin{equation}
  G_{r,\ell}(\infty;j\omega)=G^{0}_{r,\ell}(j\omega)-\frac{\beta_i(j\omega)\gamma_i(j\omega)}{\alpha_i(j\omega)} ,
  \label{eq:ceiling}
\end{equation}
\admk{the response of the network with node $i$ held fixed. This limit depends on baseline quantities alone\armk{ and is finite for $D\succ0$} (Materials and Methods). The correction in Eq.~\ref{eq:mobius_closed_form} equals $(\beta_i\gamma_i/\alpha_i)\,\varphi/(1+\varphi)$ with $\varphi=j\omega\Delta D_i\alpha_i$. Passivity gives $\mathrm{Re}\,\varphi\ge0$, which confines $\varphi/(1+\varphi)$ to the disk with diameter from $0$ to $1$. At every frequency, the gain therefore stays within the disk whose diameter joins $G^{0}_{r,\ell}$ to $G_{r,\ell}(\infty)$. Damping at node $i$ can thus change a channel by at most $|\beta_i\gamma_i/\alpha_i|$. \ajmk{The \armk{disk}} uses no property of the network beyond passivity. The same baseline quantities decide the direction of the change. With enough damping, node $i$ raises the response of a channel at a given frequency if the channel responds more strongly with node $i$ held fixed, $|G_{r,\ell}(\infty)|>|G^{0}_{r,\ell}|$. Otherwise it lowers the response (\akmk{Eq.~\ref{eq:direction}}). In the three units, the peak of the channel rises toward the response with unit 1 held fixed as $\Delta D$ grows (Fig.~\ref{fig:minimal}C and section S2).}

\admk{The limit of Eq.~\ref{eq:ceiling} explains why a weakly responding channel can open far. The opening ceiling $\chi_i$ is the opening ratio in this limit. It equals the peak of $|G^{0}_{r,\ell}-\beta_i\gamma_i/\alpha_i|$ divided by the baseline peak. Because the gain stays within the disk, every finite increment gives $\rho\le1+\chi_i$.\armk{ A finite increment can exceed $\chi_i$, but not $1+\chi_i$.} On the \armk{$\ell_{4,10}$} channel the opening ratio rises monotonically with the damping at node \armk{4} toward its ceiling of \armk{$27$} (Fig.~\ref{fig:local-damping-fragility}D). On a latent channel $|G^{0}_{r,\ell}|\ll|\beta_i\gamma_i/\alpha_i|$, so the numerator of $\chi_i$ approaches the peak of $|\beta_i\gamma_i/\alpha_i|$. \afmk{This peak need not fall with the baseline response}, as measured by $\nu$. A large opening therefore requires that $|\beta_i\gamma_i/\alpha_i|$ not vanish together with $G^{0}$. At nodes \armk{4 and 10} it does not. A force applied at node \armk{4 or 10} itself drives the pair unequally, so \armk{$\beta_4$ and $\beta_{10}$} are not suppressed. They are two orders of magnitude larger than at any other node. Across the random networks, the ceiling relative to the strongest baseline response, $\chi\nu$, falls far more slowly than $\nu$ for weakly responding channels (Fig.~\ref{fig:ensemble}D).}%

\admk{A large opening also requires the two nodes of the cell to be nearly matched. For the channel from node \armk{5} to \armk{$\ell_{4,10}$} in the ten-node network of Fig.~\ref{fig:ensemble}A, the median opening ratio falls as the parameter spread grows. With a log-normal spread \afmk{$\sigma$} of $5\%$ in the coupling or in the inertia, it falls to about \armk{$1.4$}. The same spread in the damping leaves it at \armk{$14$} (Fig.~\ref{fig:ensemble}, E to G). Coupling and inertia set the frequencies of the two nodes, so a mismatch in either raises the baseline peak strongly. The damping enters the dynamic stiffness only through $j\omega D$, which is small against $\omega^2 M$ and $L$ at the resonance. The same mismatch in the damping therefore raises the baseline peak \afmk{much less} (section S3). The spread acts through the mismatch it creates between the two nodes.%
 \atmk{The baseline peak grows roughly in proportion to $\sigma$, whereas the median opened peak changes far less (section S3)\aymk{.}} Across the random ensembles, the share of \aomk{networks with a channel} opened by more than a factor of $10$ likewise falls as the uniform spread of all parameters grows (table~\ref{tab:spread_scan}). Large openings nevertheless do not require nodes matched by design. At a uniform spread of $\pm10\%$, about one in five networks of $80$ nodes still contains a channel opened by more than a factor of $10$ (table~\ref{tab:spread_scan}).}%

\begin{figure*}[!t]
  \centering
  \includegraphics[width=492.2bp,trim=9bp 9.5bp 9.3bp 4.3bp,clip]{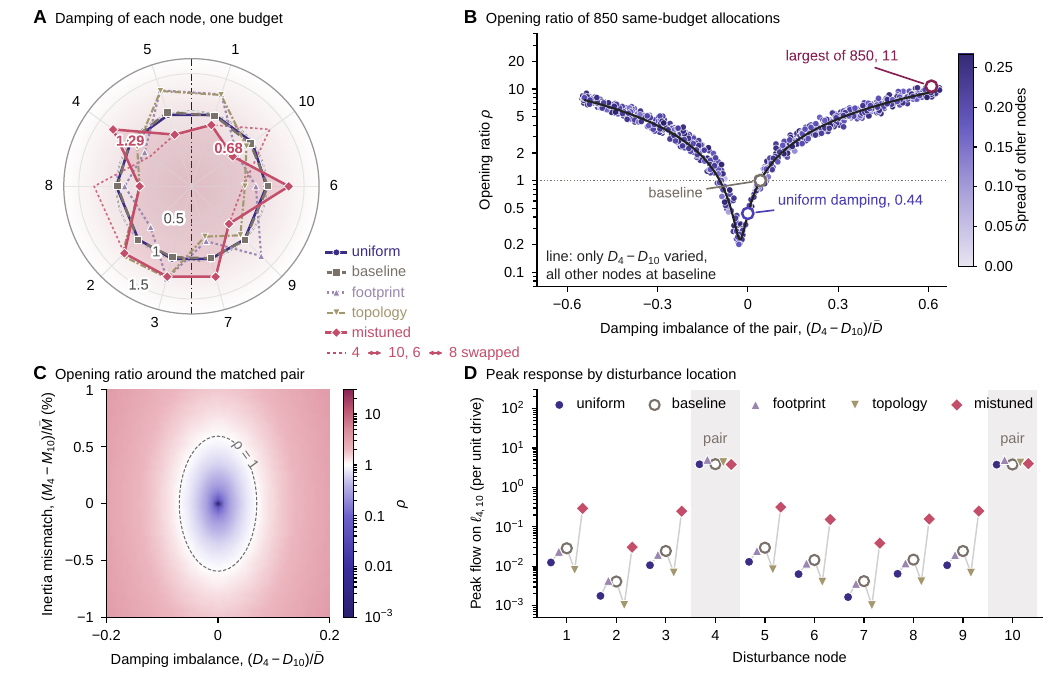}%
  \caption{\textbf{Damping placement under a fixed budget.}
  \acmk{Ten-node network of \aemk{Fig.~\ref{fig:ensemble}A}, with the disturbance at node~\armk{5}. Damping is reallocated under the same total budget and device bounds with inertia fixed.
  (\textbf{A}) Damping of each node relative to the network mean for five allocations (Materials and Methods), with the \admk{identically positioned} pairs \armk{4, 10 and 6, 8} placed symmetrically about the vertical axis. The dotted line shows the mistuned allocation with the dampings exchanged within each pair, and numbers give the dampings of nodes \bdmk{4 and 10} in the mistuned allocation.
  (\textbf{B}) Opening ratio $\rho$ of the \armk{$\ell_{4,10}$} channel for $850$ allocations of the same budget, versus the damping imbalance of the \admk{identically positioned} pair, \armk{$(D_4-D_{10})/\bar D$}. Color shows the spread of the damping of the other eight nodes from baseline, the Euclidean norm of their changes relative to that of their baseline values. The line shows \armk{$D_4-D_{10}$} varied alone, with all other nodes at baseline. Open circles mark the baseline, uniform damping and the largest of the $850$ allocations, and the dotted line marks $\rho=1$.
  (\textbf{C}) $\rho$ with uniform damping at the other eight nodes, versus the damping imbalance %
 and the inertia mismatch of nodes \armk{4 and 10}, with their couplings to node~\armk{1} made equal. The line marks $\rho=1$.
  (\textbf{D}) Peak flow on \armk{$\ell_{4,10}$} for a disturbance at each node under the five allocations of (A). The gray \armk{bands mark} disturbances on the pair.}}
  \label{fig:channel_aligned_reallocation}
\end{figure*}

\subsection*{\cmk{\emk{Placement decides the response}}}

\admk{The same damping budget can open a latent channel or keep it latent. The channel opens in proportion to the damping difference between nodes \armk{4 and 10}. It stays latent when the two are damped alike, whatever the other eight nodes receive, as the partition requires. In the ten-node network, the damping is reallocated among the nodes with the total budget and the device bounds held fixed (Materials and Methods). Across random allocations, the opening ratio $\rho$ of the \armk{$\ell_{4,10}$} channel spans \bdmk{nearly} \armk{two orders} of magnitude and follows the damping imbalance of nodes \armk{4 and 10} (Fig.~\ref{fig:channel_aligned_reallocation}B). The allocation that opens the channel most (the mistuned allocation) raises it by a factor of \armk{$11$}. Exchanging the dampings within each pair of identically positioned nodes \apmk{leaves this opening nearly unchanged}, as the exchange symmetry of nodes \armk{4 and 10} suggests (Fig.~\ref{fig:channel_aligned_reallocation}A \anmk{and sections S2 and S8}). Uniform damping and the footprint- and topology-based allocations lower the channel by \armk{$18\%$ to $72\%$}, because they give nodes \armk{4 and 10} nearly equal damping. Damping alone cannot close the channel, because the inertia and couplings of nodes \armk{4 and 10} differ slightly.\atmk{ Keeping it latent requires the two nodes to stay closely matched} (Fig.~\ref{fig:channel_aligned_reallocation}C). If instead an increment of $\varepsilon=10\%$ of the total damping is added to the baseline and spread at random over the nodes, the channel opens in \armk{$56\%$} of the allocations (fig.~\ref{fig:randdir}). For disturbances entering node \armk{4 or 10, $\ell_{4,10}$} carries the strongest flow of the network. None of the five allocations of Fig.~\ref{fig:channel_aligned_reallocation}A then changes this response by more than \armk{$31\%$} (Fig.~\ref{fig:channel_aligned_reallocation}D). The placement therefore has to be chosen for the channel and disturbance of concern. \afmk{A placement that keeps such channels latent damps the nodes of a cell alike, for instance identical converters at one bus, and Eq.~\ref{eq:mobius_closed_form} shows before the intervention which other placements open which channels and by how much.}}%

\section*{Discussion}

\aymk{We have shown that a passive increase of damping can open a latent channel by more than an order of magnitude while the smallest damping ratio rises and the energy stored in the network under random disturbances falls (section S4.3).} \aymk{}%
\awmk{The opening has a simple origin. \bamk{A disturbance that enters the network is shared among the nodes, and a change in the damping of one node changes its share at every frequency.} Two \aymk{interchangeable} nodes take equal shares, so the branch between them carries no flow. Damping one of them makes the shares unequal, and the difference flows through that branch. Near the antiphase resonance of the pair, the difference is amplified\azmk{. The disturbance could not excite this resonance before, but the added damping mixes it into a nearby resonance that the disturbance does excite.} \azmk{Damping} thus \aymk{shifts} part of \aymk{the} response onto a branch that was idle. This picture also tells when to expect the effect: two nearly \aymk{interchangeable} nodes, a disturbance that enters elsewhere, and a change that \aymk{leaves them unequal, however much is added elsewhere in the network}.\aymk{ The same increment can also raise channels that already respond strongly, including the strongest flow of the network.} Before and after the intervention, the stored energy can only decay (section S4.3), so the amplification is a change of the steady forced response, not the transient growth of non-normal networks}\cite{Trefethen1993,Asllani2018}\awmk{.}

\aymk{Modal screening ranks channels by the residue of a mode at baseline and therefore places the most weakly responding channels last (fig.~\ref{fig:competing-indices}). \azmk{Such a ranking \bdmk{overlooks} the oscillations that added damping can raise on these channels.} The opening is instead a change of that residue produced by the added damping. In the closed form (Eq.~\ref{eq:mobius_closed_form}), this change is set by how strongly the input reaches the damped node and how strongly that node reaches the output. The response of the network before the change thus tells which placement opens which channel and how far, and for any added damping\aymk{, however large,} it bounds the opening (Eq.~\ref{eq:ceiling}).}%

\axmk{Three lines of work come close to this one. Symmetry, and more generally an equitable partition of the nodes, can hide a mode from an input or an output, and breaking it restores controllability and observability}\cite{RubinMeadows1972,Rahmani2009,Martini2010,OClery2013,Whalen2015}\axmk{\aymk{.} \aymk{T}he same partitions \aymk{also} underlie cluster synchronization}\cite{Pecora2014,Schaub2016}\axmk{. In the damped wave equation on a graph, damping at a node where a mode vanishes leaves that mode undamped}\cite{Caputo2013}\axmk{\aymk{.} \aymk{F}or two nodes with the same other neighbors, the antiphase mode of the pair is such a mode, as it vanishes at their common neighbors. These results decide whether a mode reaches a channel, not how far the channel rises. Damping can also raise forced responses in mechanical structures. In mistuned bladed disks, damping that differs between blades raises the forced response by up to about three times the tuned response, with the mean damping fixed}\cite{Khemiri2013}\axmk{, and in a tuned absorber damping fills the antiresonance}\cite{Ormondroyd1928,DenHartog1956}\axmk{\aymk{.} \aymk{I}n both cases the symmetry or the tuning is designed, and the response is strong before the change. Rank-one modification theory gives the forced response after a single element is changed and is used to design such changes}\cite{MottersheadRam2006,GhandchiTehrani2006}\axmk{. Here these lines meet: a change that only dissipates energy opens channels that are nearly silent, the near-symmetry that silences them arises by itself among \aymk{identically positioned nodes}, and the rank-one formula, applied to such a channel, predicts the opening and bounds it from baseline quantities.}%

\aymk{While we focused on damping, because it is the only adjustable parameter that can only absorb energy and its effect is therefore the least expected, other local changes open latent channels as well. Lowering the inertia of a single node leaves the spectrum nearly unchanged and opens them more often and by larger factors than added damping in the random networks (section S12), and it also opens them on the pan-European grid (fig.~\ref{fig:real-grid}E and section \bcmk{S12}). A change of inertia is not bounded by passivity, and it can raise channels that carry less than a tenth of the strongest flow under the same disturbance to nearly that flow or above it, whereas added damping raises them to at most about a quarter of it (section \bcmk{S12}). Lowering the coupling of a single line acts in the same way (section \bcmk{S12}). \azmk{The closed form contains} nothing specific to power grids\azmk{, the ceiling holds wherever the damped node is passive}, and latent channels open in all the oscillator models of Fig.~\ref{fig:universality}.}%

\aymk{These conclusions are based on linearized dynamics. In the full swing equation, the channel still opens well beyond the range in which the linearized prediction is accurate\azmk{. When the branch carries a large steady flow, the opening falls with the amplitude, from a factor of $11$ to between $4.5$ and $6.2$ at the largest excursion} (section S11). Beyond the linear range, an opened channel can also switch a bistable unit, of the kind used to model epilepsy}\cite{KatoKori2024}\aymk{, into a sustained oscillation that remains after the disturbance has ended, while every eigenvalue of the rest state moves away from the imaginary axis (fig.~\ref{fig:bistable} and section S10). The results on the pan-European grid rest on a model that represents primary control and load response by a constant nodal damping\azmk{, and n}o measured response is used (sections S6 and S9).}\admk{ The resonances studied here lie in the electromechanical band\aymk{.} In this band, \aymk{the} droop and virtual synchronous machine controls \aymk{of inverter-based generators} reduce to swing equations with virtual inertia and damping\cite{DArcoSuul2014}. The mechanism and the predictor need only a linearized model with an input port and a monitored output, so the same framework applies to more detailed models of these units. \agmk{Which channels are latent will change with the model.}}

\aymk{These results bear on networks built from similar units. }\admk{In mistuned bladed disks, the amplification of the forced response has been linked to shortened component life\cite{JonesCross2003}. \aymk{In power grids, h}omogenizing generator parameters can strengthen the stability of the synchronous state, and suitable heterogeneity can strengthen it further\cite{Molnar2020,Molnar2021}. For forced responses\aymk{, in contrast,} near-uniform parameters make latent channels very weak at baseline, and a local change of damping can open them. \aymk{A large opening requires the two nodes to be closely matched in \azmk{the ratio of coupling to inertia} (Fig.~\ref{fig:ensemble}, E and F).} %
 Synchronous generators of different type and rating\cite{Kundur1994} rarely form such a cell, whereas \azmk{identical units at one plant or one bus do, }\azmk{whether}\azmk{ synchronous (section S3) or inverter-based}\azmk{. For a disturbance entering elsewhere, the} difference of their outputs is then zero at baseline, and it grows in proportion to the change when the damping or the inertia setting of one unit is changed. For power grids with many distributed generators, our results suggest that damping should be placed with the disturbance sources and the monitored outputs in mind.}

\section*{Materials and Methods}

\subsection*{\ahmk{\lmk{Externally equitable partitions}}}

\lmk{A partition of the nodes into cells is \gmk{externally} equitable\gmk{\cite{OClery2013,Schaub2016}} \ahmk{if} every node of a cell has the same total coupling $\sum_{n\in C'}(-L_{in})$ to each other cell $C'$\gmk{, whatever the connections inside the cell}. With $P$ the $N\times N_c$ indicator matrix of the $N_c$ cells, $LP=PL_\pi$ for an $N_c\times N_c$ quotient matrix $L_\pi$. If $M$ and $D$ are constant on each cell, $MP=PM_\pi$ and $DP=PD_\pi$, so $Z(s)P=PZ_\pi(s)$ and the range of $P$ is invariant under $Z(s)^{-1}$. If the disturbed node $r$ forms a cell of its own, $e_r$ lies in that range, \acmk{and} $Z(s)^{-1}e_r$ takes equal values on the nodes of each cell\acmk{. Then} $c_\ell^\top Z(s)^{-1}e_r=0$ for every branch $\ell$ inside a cell, and for every output $c$ with $c^\top P=0$, that is, every output that compares nodes of one cell. The factor $\beta_i=c_\ell^\top Z_0^{-1}e_i$ of Eq.~\ref{eq:mobius_closed_form} vanishes by the same argument whenever node $i$ forms a cell of its own in such a partition\acmk{. The channel also} stays zero when the damping is raised equally within every cell, because the partition remains equitable.}

\lmk{\afmk{The coarsest \ahmk{externally} equitable partition} with $r$ in its own cell\gmk{, the minimal partition with respect to node $r$ of ref.~\cite{OClery2013},} is obtained by refining the partition $\{\{r\},\text{rest}\}$ until every node of a cell has the same number of neighbors in each \qmk{other} cell\qmk{. This is the color refinement of Weisfeiler and Leman\cite{WeisfeilerLeman1968} with the neighbors inside a node's own cell left uncounted}, which is the condition above for the topology with equal couplings. For the ensembles of Fig.~\ref{fig:ensemble} this partition is computed on the topology of each network for every disturbed node, and a branch is a twin branch if its two ends have the same other \afmk{neighbors (section S3)}.}%

\subsection*{\ahmk{\rson{}Oscillator models\rsoff{}}}

\rson{}The models of Fig.~\ref{fig:universality} share the triangle of Fig.~\ref{fig:minimal}, with couplings \hmk{$K_{in}$} ($K_{13}=K_{23}=K=1$), a drive $p(t)$ at unit 3 and the flow on branch 1--2 as output\cmk{\acmk{. Their} equations and parameters are listed in table~\ref{tab:osc_models}}. The local dissipation parameter of unit 1 or 2 is raised from its common value by $\Delta D$, up to $5$ times that value, and each model is linearized about its stable state. \aymk{In model G, raising $\lambda_i$ also shifts the rest state, so $\Delta D$ there is a local parameter that distinguishes unit 1 from unit 2 rather than a pure dissipation (section S1.1).}

\afmk{The peak is read} from the frequency response on a dense grid from half the lowest oscillatory frequency of the baseline network upward, refined around each mode. \afmk{At every grid point} the peak lies within $2.5\%$ of an oscillatory mode frequency, and all states remain linearly stable over the scanned range. The grid has $41$ values of $\Delta D$ and $31$ of $K_{12}$ between $0.5K$ and $2K$, with eight further values of $\Delta D$ between \hmk{$0.2\%$ and $10\%$ of the common value} for the contours. Dissipation at unit 2 gives the same \akmk{opened response} as at unit 1 \apmk{over the whole grid}\aymk{.}\rsoff{}

\subsection*{\ahmk{Random network ensembles}}

The topologies of Fig.~\ref{fig:ensemble} are generated by the spatially embedded random growth model of ref.~\cite{Schultz2014}, with nodes placed uniformly in the unit square, \hmk{one initial node and the growth parameters of ref.~\cite{Nitzbon2017}\acmk{. These are} probabilities $1/5$, $3/10$ and $1/10$ of an extra line from a new node, of a line between existing nodes and of splitting a line, and a cost--redundancy exponent of $1/3$}. Linearized line couplings, inertia and damping are drawn independently and uniformly within \fon{}$\pm5\%$\foff{} of \wmk{$2.26\times10^{4}$}, \hmk{$150$ and $280$}\afmk{, so that the ratios of coupling and damping to inertia are about $151$~s$^{-2}$ and $1.87$~s$^{-1}$. This places} the median damping ratio of the oscillatory modes at \emk{$0.05$}\foff{}. These values are of the same order as those of ref.~\cite{Zhang2019} and correspond to an inertia constant of about $4$~s and a line reactance of about $0.26$~p.u. at $50$~Hz.%

\ymk{\afmk{The band} $0.2$--$10$~Hz is sampled at} $400$, $350$, $300$ and $220$ frequencies for $N=10$, $20$, $40$ and $80$\ymk{. On this grid} the closed form yields for every channel the baseline response strength, the opening ratio when the damping of a single node is raised by the network's mean damping, $\Delta D_i=\hmk{\bar D=\sum_nD_n/N}$, maximized over nodes, and the ceiling for unlimited damping. \ahmk{On a $0.01$~Hz grid, for} each channel opened by more than a factor of $2$, the post-intervention pole $\cmk{s_k}'$ nearest to $j\omega$ at the worst-case frequency and the nearest baseline pole $\cmk{s_k}$ \acmk{are identified. They split} the modal gain ratio into the \dmk{residue factor} $|R_k'|/|R_k|$ and the \dmk{pole factor} $|j\omega-\cmk{s_k}|/|j\omega-\cmk{s_k}'|$ \dmk{(Fig.~\ref{fig:ensemble}C)}.

\imk{\ahmk{Shares} of networks \afmk{have} $95\%$ binomial confidence intervals of \kmk{$\pm2$} to $\pm6$ percentage points ($n=300$ per size).}

\subsection*{\ahmk{\nmk{Random ten-node network and damping intervention}}}%

\nmk{The random ten-node network is the $N=10$ realization of Fig.~\ref{fig:ensemble}A}\aymk{ (Random network ensembles), with all parameters in table~\ref{tab:ten_machine_parameters}}\ymk{. The \emk{inertias} of nodes \armk{4 and 10} \emk{differ} by \armk{$2.4\%$} and their couplings to node~\armk{1} by \armk{$2.6\%$}\azmk{, but the ratio of coupling to inertia, which sets how fast each node oscillates against node~1, differs by only $0.14\%$\bcmk{ (computed before the rounding of table~\ref{tab:ten_machine_parameters})}}.} \xmk{\kmk{The couplings $-L_{in}$ are drawn directly and correspond to an operating point} with \hmk{$\mathrm{d}g/\mathrm{d}\theta>0$} on every edge\ymk{, so that the weighted Laplacian $L$ is positive semidefinite}\kmk{\acmk{. An} explicit operating point is constructed only for section S11}.} Stability is assessed on the spectrum of the linearized model after removal of the uniform phase-drift mode\xmk{, which has zero residue on every flow output}, and every reported configuration has all remaining eigenvalues in the open left half-plane. The disturbance is a unit sinusoidal power injection at \nmk{node \armk{5}}. The monitored outputs are the linearized flows \wmk{\hmk{$c_\ell^\top\theta$}} on all \nmk{\armk{$11$} branches}. A single-node intervention adds \nmk{the network's mean damping, $\varepsilon=10\%$ of its total damping,} to one node and leaves inertia and all other parameters unchanged, which raises the damping of \nmk{node \armk{4 from $280$ to $557$}\acmk{.}}

For each channel the peak gain $\Gamma$ \atmk{of Eq.~\ref{eq:opening_ratio}} and the worst-case frequency $\omega^\star$ are read on a dense grid over the band \afmk{$\mathcal{W}$ of \armk{$0.2\le\omega/2\pi\le10$}~Hz,} whose lower edge lies \nmk{below the lowest mode at \armk{$0.884$}~Hz}. Opening ratios compare each channel with its own baseline peak gain $\Gamma_0$ and equal the corresponding ratios of root-mean-square responses to a sinusoidal drive. \nmk{In the random ten-node network}, the resonant pole lies at \nmk{\armk{$3.373$}~Hz and the baseline worst-case frequency at \armk{$3.398$}~Hz}. \revon{}\afmk{The baseline response strength $\nu$ is computed over the same band}\aymk{.}\revoff{}

\fon{}\ahmk{For} Fig.~\ref{fig:ensemble}, E to \almk{G}, the couplings, the \ahmk{inertias} or the dampings of the ten-node network of Fig.~\ref{fig:ensemble}A \acmk{start} from identical values\acmk{. They} are multiplied one class at a time by \hmk{independent log-normal factors of mean $1$ and logarithmic standard deviation $\sigma$}, which preserve their expected values\ymk{. The $3000$ values of $\sigma$ are} drawn log-uniformly between $0.01\%$ and $15\%$, of which those above $0.1\%$ are shown. \nmk{The channel runs from node \armk{5} to the branch between nodes \armk{4 and 10}, which have the same \akmk{other} neighbors, and the damping of node \armk{4} is raised by the mean damping.}\foff{}

\subsection*{\xmk{Closed-form predictor}}

The main text uses the second-order form $Z(s)=s^2M+sD+L$. Numerical computations use the equivalent first-order realization with state \hmk{$x=(\theta,\dot\theta)$}, which yields the same poles, zeros and channel gains to machine precision. \ymk{Adding damping $\Delta D_i>0$ at a single node $i$ is a rank-one modification of the dynamic stiffness matrix,}
\begin{equation}
  Z(\Delta D_i;j\omega)=Z_0(j\omega)+j\omega\,\Delta D_i\,e_ie_i^\top ,
  \label{eq:rankone_update}
\end{equation}
\ymk{where $Z_0$ is the baseline dynamic stiffness. The added element dissipates the instantaneous power $\Delta D_i\dot\theta_i^{\,2}\ge 0$, so the modification only absorbs energy.} \ahmk{\bdmk{The disk uses no property of the network other than its passivity at the damped node, $\mathrm{Re}(j\omega\alpha_i)\ge0$.} Positive realness keeps the denominator of Eq.~\ref{eq:mobius_closed_form} from vanishing anywhere on $\Delta D_i\in(0,\infty)$.} \armk{Positive realness alone does not exclude $\alpha_i=0$. }\admk{With $D\succ 0$, $\mathrm{Re}(j\omega\alpha_i)>0$ at every $\omega\neq0$, so $\alpha_i$ has no zero on the real frequency axis and the limit of Eq.~\ref{eq:ceiling} is finite at every frequency.} \afmk{The disk of the main text follows from $\varphi/(1+\varphi)-\tfrac12=(\varphi-1)/[2(\varphi+1)]$, since $|\varphi-1|\le|\varphi+1|$ whenever $\mathrm{Re}\,\varphi\ge0$.} Scanning all $N$ candidate locations costs one baseline receptance evaluation per frequency and $O(N)$ scalar operations instead of $N$ matrix inversions, so \aymk{all} combinations of damping node and branch follow before any intervention is made\aymk{.} \dmk{The closed form agrees with a direct re-solution to machine precision for both the ten-node network and the pan-European grid. \aymk{Section S1 lists the model equations and parameters, and sections S2 and S4 give the extended derivations.}}

\aymk{Changes at several nodes or branches add to $Z_0$ a term $\Delta Z=U\Lambda U^\top$. Each column of $U$ is $e_i$ for a change at node $i$ or $e_i-e_j$ for a change of branch $i$--$j$, and the corresponding diagonal entry of $\Lambda$ is $j\omega\Delta D_i$, $-\omega^2\Delta M_i$ or $\Delta K_{ij}=-\Delta L_{ij}$, the change of the coupling of that branch. For an input vector $b$, applying the inversion behind Eq.~\ref{eq:mobius_closed_form} to this change of higher rank gives}
{%
\begin{equation}
G=c_\ell^\top Z_0^{-1}b-c_\ell^\top Z_0^{-1}U\bigl(I+\Lambda\,U^\top Z_0^{-1}U\bigr)^{-1}\Lambda\,U^\top Z_0^{-1}b ,
\label{eq:woodbury}
\end{equation}}
\aymk{which reduces to Eq.~\ref{eq:mobius_closed_form} for damping at a single node and can be checked by multiplication with $Z$. Because $M$, $D$ and $L$ are real and symmetric, the imaginary part of $x^{H}Z(j\omega)x$ is $\omega\,x^{H}Dx$, so $Z(j\omega)$ is nonsingular at every $\omega\neq0$ as long as the damping matrix remains positive definite. The matrix in parentheses is then nonsingular as well, since its determinant equals $\det Z/\det Z_0$.}%

\jmk{Writing Eq.~\ref{eq:mobius_closed_form} as $G_{r,\ell}(\Delta D_i)=\big(G^{0}_{r,\ell}+\Delta D_i\,q_i\,G_{r,\ell}(\infty)\big)/(1+\Delta D_i\,q_i)$ with $q_i=j\omega\alpha_i$, one finds $|G_{r,\ell}(\Delta D_i)|>|G^{0}_{r,\ell}|$ at frequency $\omega$ if and only if}
{\color{black}%
\begin{equation}
  \big|G^{0}_{r,\ell}+\Delta D_i\,q_i\,G_{r,\ell}(\infty)\big|>\big|G^{0}_{r,\ell}\big|\,\big|1+\Delta D_i\,q_i\big| .
  \label{eq:direction}
\end{equation}}

\subsection*{\ron{}Same-budget allocations\roff{}}

Damping is reallocated with the total fixed, \hmk{$\sum_nD_n=D_{\mathrm{tot}}$}, and each \dmk{node} bounded within $\pm30\%$ of its baseline damping, while inertia stays at its baseline distribution. The $850$ allocations of Fig.~\ref{fig:channel_aligned_reallocation}B multiply the baseline damping by \hmk{independent log-normal factors of median $1$ and logarithmic standard deviation $0.45$} and are projected onto these constraints\acmk{. All} are linearly stable because every damping stays positive. \afmk{The uniform allocation has equal damping at every node}, \ymk{and} the mistuned \afmk{allocation} is the \afmk{one} with the largest opening among the $850$\ymk{. The} footprint-based and topology-based \afmk{allocations} distribute the budget in proportion to the squared velocity amplitude of each \nmk{node} at the baseline worst-case frequency \afmk{plus $5\%$ of its mean} and to the weighted degree $L_{ii}$ of \ahmk{each node}, each projected onto the same constraints. \nmk{For Fig.~\ref{fig:channel_aligned_reallocation}C, the inertia of nodes \armk{4 and 10} and their couplings to node~\armk{1} are set to their means\acmk{. An} inertia mismatch is then imposed with opposite signs at the two nodes, and the damping imbalance is added to uniform damping, on a $121\times121$ grid.}

\subsection*{Time-domain responses}

\mkon{}\xmk{The traces of Fig.~\ref{fig:minimal}B are responses of the linearized three-unit network} with \aumk{the parameters of Fig.~\ref{fig:minimal}, $m=1$, $K_{12}=K_{13}=K_{23}=K=1$ and $d=0.04$,} \xmk{to a unit sinusoidal drive at unit 3 at} \aumk{$\omega_0=\sqrt3$}\xmk{\ymk{. They start} from rest and \ymk{are} propagated exactly with the matrix exponential of the model augmented by the drive oscillator}. \smk{The peaks of Fig.~\ref{fig:minimal}C are \aemk{read over the band defined in Oscillator models, from half the lowest oscillatory frequency of the baseline network upward}.}\mkoff{} \nmk{The traces of Fig.~\ref{fig:local-damping-fragility}A are responses of the \xmk{linearized ten-node network} to band-limited white noise over $0.2$--$10$~Hz injected at node \armk{5}, with one noise realization for all dampings\ymk{. Distributions} are taken over $30$~s\nmk{, and rms values follow exactly from the frequency response over the band}.} \hmk{\ahmk{Section S11 repeats this test in the full swing equation.}}

\bibliographystyle{sciencemag}
\bibliography{references}

\section*{Acknowledgments}
\aymk{We thank Adilson E. Motter for comments on the manuscript and for suggesting the analysis of the absolute size of the openings, a consistent screening of the pan-European grid, a sharper statement of the bounds and the comparison with damping mistuning, symmetry-induced unobservability and rank-one modification.}%

\subsection*{Supplementary Materials}
Supplementary Methods\\
Supplementary Text\\
\rson{}Figs. S1 to \rsoff{}\lmk{S6}\\
Tables S1 to \anmk{S7}\\
\bcmk{References (56--61)}
\newpage
\onecolumn
\raggedbottom%
\renewcommand{\arraystretch}{1.15}
\gdef\thefigure{S\arabic{figure}}
\gdef\thetable{S\arabic{table}}
\gdef\theequation{S\arabic{equation}}
\gdef\thepage{S\arabic{page}}
\setcounter{figure}{0}
\setcounter{table}{0}
\setcounter{equation}{0}
\setcounter{page}{1}

\begin{center}
{\sffamily\large\bfseries Supplementary Materials for\par}
\vspace{0.4em}
{\sffamily\LARGE\bfseries\color{saRed}\scititle\par}
\vspace{0.8em}
{\sffamily\normalsize \acmk{\sciauthors}\par}
\vspace{0.3em}
{\sffamily\footnotesize \acmk{\sciaffils}\par}
\vspace{0.8em}
\end{center}

\subsubsection*{This PDF file includes:}
Supplementary Methods (section S1)\\
Supplementary Text (sections S2 to \jmk{S12})\\
\rson{}Figs. S1 to \rsoff{}\lmk{S6}\\
Tables S1 to \anmk{S7}\\
\bcmk{References (56--61)}
\vspace{1em}

\FloatBarrier%
\subsection*{Section S1. Supplementary methods}

\subsubsection*{\cmk{S1.1 Oscillator models}}
\cmk{Table~\ref{tab:osc_models} lists the eight models of Fig.~\ref{fig:universality}. \ahmk{Network, drive and output are those of Materials and Methods (Oscillator models).}}\aymk{ Models A to C share the injection of $0.4$ into the lines at unit 3. In model B the line resistance equals the reactance, a strongly lossy case in which $17\%$ of the $0.4$ injected by unit 3 is lost on the lines. In model C the voltages $V_i$ stay between $1.18$ and $1.41$, and the voltage dynamics couple to the rotor angles only through the line loading and add about $1\%$ to the synchronizing power, so its opening follows that of model A. In model D the ground springs are ten times stiffer than the couplings, as in structures with stiff supports and weak connections. Raising $\lambda_i$ in model G also lowers the rest state and the frequency of the unit, which stops oscillating beyond $\lambda_i\approx1.86\,\lambda$.}

\begin{table}[!htbp]
\centering
\caption{\cmk{\textbf{Oscillator models of Fig.~\ref{fig:universality}.} $\Delta D$ raises the listed local parameter of unit 1 (or 2) from its common value. Each model is linearized about its stable state.}}
\label{tab:osc_models}
\vspace{0.4em}
{\footnotesize\cmkcolor
\renewcommand{\arraystretch}{1.35}
\newcommand{\modelsep}{\noalign{\vskip1.5pt{\color{black!35}\hrule height0.3pt}\vskip1.5pt}}%
\newcommand{\vc}[1]{\begin{tabular}[c]{@{}l@{}}#1\end{tabular}}%
\begin{tabular}{@{}l l l l l@{}}
\hline
 & Model & Equations of unit $i$ and output $F_{12}$ & $\Delta D$ on & Parameters \\
\hline
A & Second-order Kuramoto\cite{Filatrella2008} & \vc{\hmk{$\ddot\theta_i+D_i\dot\theta_i=\Omega_i+\sum_n K_{in}\sin(\theta_n-\theta_i)+p_i$}\\ $F_{12}=K_{12}\sin(\theta_1-\theta_2)$} & $D_i$ & \hmk{$D_i=0.04$}, \hmk{$\Omega=(-0.2,-0.2,0.4)$} \\
\modelsep
B & \vc{Second-order Kuramoto,\\ lossy lines\cite{Kundur1994}} & \vc{\hmk{$\ddot\theta_i+D_i\dot\theta_i=\Omega_i-\sum_n K_{in}\big[\sin(\theta_i-\theta_n)+1-\cos(\theta_i-\theta_n)\big]+p_i$}\\ $F_{12}=K_{12}\big[\sin(\theta_1-\theta_2)+1-\cos(\theta_1-\theta_2)\big]$} & $D_i$ & \vc{\hmk{$D_i=0.04$, $\Omega_3=0.4$,}\\ \hmk{$\Omega_1=\Omega_2=-0.166$;}\\ \hmk{line resistance equal to reactance}} \\
\modelsep
C & Third-order generator\cite{Schmietendorf2014} & \vc{\hmk{$\ddot\theta_i+D_i\dot\theta_i=\Omega_i-\sum_n \tfrac12K_{in}V_iV_n\sin(\theta_i-\theta_n)+p_i$}\\ \hmk{$T\dot V_i=V_{\mathrm f}-V_i+X\sum_n \tfrac12K_{in}V_n\cos(\theta_n-\theta_i)$}\\ \hmk{$F_{12}=\tfrac12K_{12}V_1V_2\sin(\theta_1-\theta_2)$}} & $D_i$ & \vc{as A\\ $T=2$, $X=0.2$, \hmk{$V_{\mathrm f}=1$}} \\
\modelsep
D & Spring--mass network\cite{Montanari2026} & \vc{\hmk{$\ddot\theta_i+D_i\dot\theta_i+K_{\mathrm g}\theta_i=\sum_n K_{in}(\theta_n-\theta_i)+p_i$}\\ \hmk{$F_{12}=K_{12}(\theta_1-\theta_2)$}} & $D_i$ & \hmk{$D_i=0.04$}, \hmk{$K_{\mathrm g}=10$} \\
\hline
E & Stuart--Landau\cite{Pikovsky2001} & \vc{\hmk{$\dot z_i=(\mu_i+j\omega_{\mathrm{nat}})z_i-|z_i|^2z_i+\xi\sum_n K_{in}(z_n-z_i)+p_i$}\\ $F_{12}=\xi K_{12}\,\mathrm{Re}(z_1-z_2)$} & \hmk{$|\mu_i|$} & \vc{\hmk{$\mu=-0.05$}, \hmk{$\omega_{\mathrm{nat}}=1$}, $\xi=0.05$\\ $p_i$ real} \\
\modelsep
F & van der Pol\cite{Pikovsky2001} & \vc{\hmk{$\ddot\theta_i-\mu_i(1-\theta_i^2)\dot\theta_i+\theta_i=\xi\sum_n K_{in}(\theta_n-\theta_i)+p_i$}\\ \hmk{$F_{12}=\xi K_{12}(\theta_1-\theta_2)$}} & $|\mu_i|$ & $\mu=-0.05$, $\xi=0.2$ \\
\modelsep
G & FitzHugh--Nagumo\cite{FitzHugh1961} & \vc{\hmk{$\dot\theta_i=\theta_i-\theta_i^3/3-\psi_i+\xi\sum_n K_{in}(\theta_n-\theta_i)+p_i$}\\ \hmk{$\tau\dot\psi_i=\theta_i+a-\lambda_i\psi_i$}\\ \hmk{$F_{12}=\xi K_{12}(\theta_1-\theta_2)$}} & \hmk{$\lambda_i$} & \vc{$a=0.45$, \hmk{$\tau=12.5$}\\ $\xi=0.02$, \hmk{$\lambda=0.8$}} \\
\modelsep
H & Josephson junctions\cite{Montanari2026} & \vc{\hmk{$\ddot\theta_i+D_i\dot\theta_i+J_{\mathrm c}\sin\theta_i=J_{\mathrm b}+\xi\sum_n K_{in}(\dot\theta_n-\dot\theta_i)+p_i$}\\ \hmk{$F_{12}=\xi K_{12}(\dot\theta_1-\dot\theta_2)$}} & $D_i$ & \vc{\hmk{$D_i=0.04$}, $\xi=0.05$\\ \hmk{$J_{\mathrm c}=1.5$, $J_{\mathrm b}=2J_{\mathrm c}/3$}} \\
\hline
\end{tabular}}
\end{table}

\subsubsection*{\cmk{S1.2} \nmk{Ten-node network}}
\nmk{Table~\ref{tab:ten_machine_parameters} lists all parameters of the random ten-node network of Fig.~\ref{fig:ensemble}A, used in Figs.~\ref{fig:local-damping-fragility} and \ref{fig:channel_aligned_reallocation}.}

\begin{table}[!htbp]
\centering
\caption{\nmk{\textbf{Parameters of the random ten-node network.} Realization of the random growth model with inertia, damping and coupling within $\pm5\%$ of $150$, $280$ and $2.26\times10^{4}$. Nodes \armk{4 and 10} are linked to each other and to node \armk{1} only.}}
\label{tab:ten_machine_parameters}
\vspace{0.4em}
\begin{tabular}{lcc@{\hspace{2.5em}}lcc}%
\hline
Node & Inertia $M_i$ & Damping $D_i$ & Node & Inertia $M_i$ & Damping $D_i$ \\
\hline
1 & \armk{$153.7$} & \armk{$274.3$} & 6 & \armk{$150.8$} & \armk{$281.7$} \\
2 & \armk{$147.5$} & \armk{$280.1$} & 7 & \armk{$153.5$} & \armk{$274.8$} \\
3 & \armk{$157.3$} & \armk{$270.9$} & 8 & \armk{$150.5$} & \armk{$272.9$} \\
4 & \armk{$146.3$} & \armk{$280.1$} & 9 & \armk{$151.1$} & \armk{$279.9$} \\
5 & \armk{$156.2$} & \armk{$286.2$} & 10 & \armk{$149.9$} & \armk{$268.6$} \\
\hline
\end{tabular}
\vspace{0.6em}

{\newcommand{\vc}[1]{\begin{tabular}[c]{@{}l@{}}#1\end{tabular}}%
\begin{tabular}{ll}
\hline
\emk{Line couplings \hmk{$-L_{in}$}} & \vc{\armk{$\ell_{1,3}$ $22{,}261$; $\ell_{1,4}$ $22{,}482$; $\ell_{1,5}$ $21{,}875$; $\ell_{1,10}$ $23{,}076$; $\ell_{2,3}$ $23{,}674$; $\ell_{2,7}$ $22{,}219$;}\\ \armk{$\ell_{2,9}$ $23{,}175$; $\ell_{3,9}$ $23{,}515$; $\ell_{4,10}$ $22{,}794$; $\ell_{5,6}$ $22{,}597$; $\ell_{5,8}$ $22{,}845$}} \\
Electromechanical modes & \armk{$0.884$, $1.380$, $1.949$, $2.062$, $3.373$, $3.417$, $3.673$, $3.973$, $4.496$}~Hz \\
Disturbance port $b_r$ & power injection at node \armk{5} \\
Monitored output $c_\ell$ & linearized flow on branch \armk{$\ell_{4,10}$} \\
\hline
\end{tabular}}
\end{table}

\subsection*{Section \umk{S2. \aimk{Minimal model and exchange parity}}}
\mkon{}The three-unit network of Fig.~\ref{fig:minimal} obeys $M\ddot\theta+D\dot\theta+L\theta=\hmk{b_r}\,p(t)$ with $M=mI$, $D=dI$\emk{, where $m$ and $d$ are the inertia and damping of each unit \hmk{as in Fig.~\ref{fig:minimal}}}, the Laplacian $L$ of the triangle with $K_{13}=K_{23}=K$ and $K_{12}$, \hmk{the drive $b_r=e_3$ and the monitored flow $c_\ell=K_{12}(e_1-e_2)$}. \aumk{Figure~\ref{fig:minimal} uses $K_{12}=K$, at which the antiphase and in-phase modes below coincide at $\Lambda=3K$, and Fig.~\ref{fig:universality} varies $K_{12}$.} With uniform damping the mode vectors are those of $L$: the zero mode \hmk{$\mathbf{1}/\sqrt3$ with $\mathbf{1}=(1,1,1)$}, the antiphase mode \hmk{$v_-=(1,-1,0)/\sqrt2$ with $\Lambda_-=K+2K_{12}$}, and the in-phase mode \hmk{$v_+=(1,1,-2)/\sqrt6$ with $\Lambda_+=3K$}.\aumk{ Their frequencies are $\omega_\pm=\sqrt{\Lambda_\pm/m}$, and $s_\pm$ denote the corresponding poles.} The drive \cmk{at unit 3 does not excite \hmk{$v_-$}}, \hmk{$b_r^\top v_-=0$}, and in-phase motion carries no flow on branch 1--2, \hmk{$c_\ell^\top\mathbf{1}=c_\ell^\top v_+=0$}. Every residue of the channel therefore vanishes and $G^{0}(s)\equiv0$.

\aumk{For $K_{12}=K$ the eigenspace of $\Lambda=3K$ is two-dimensional. Damping added at unit 1, $D\to D+\Delta D\,e_1e_1^\top$, acts in it only on the combination $w=(2,-1,-1)/\sqrt6$, which carries all the motion of unit 1, and leaves $u=(0,1,-1)/\sqrt2$, which does not move unit 1, undamped. For $d,\Delta D\ll\sqrt{Km}$ the two resonances near $\omega_0=\sqrt{3K/m}$ decay at the rates $d/(2m)$ for $u$ and $(d+2\Delta D/3)/(2m)$ for $w$, so the spectral margin is unchanged. The drive and the output on branch 1--2 give the two resonances opposite weights, $(c_\ell^\top u)(u^\top b_r)=-(c_\ell^\top w)(w^\top b_r)=K/2$, so that they cancel on this branch only while their decay rates are equal. At $\omega_0$,}
\begin{equation}
\aumk{|G_{r,\ell}(j\omega_0)|=\frac{K}{\omega_0 d}\,\frac{\Delta D}{3d+2\Delta D},}
\label{eq:sup_degenerate}
\end{equation}
\aumk{where $K/(\omega_0 d)$ is the largest force with identical units, carried by branches 1--3 and 2--3. The channel is linear in $\Delta D$ for $\Delta D\ll d$, reaches $20\%$ of this force at $\Delta D=d$ (\bcmk{$|G_{r,\ell}|=2.887$}, against $2.886$ exact) and tends to $50\%$ at $\omega_0$ for large $\Delta D$. Only $u$ reaches branch 2--3, whose peak therefore stays at $K/(\omega_0 d)$ for every $\Delta D$, whereas on branch 1--3 the two resonances add and the peak first falls. At this crossing the poles split at first order in $|D_1-D_2|$, whereas the channel remains linear in $D_1-D_2$, which is why the opening of Fig.~\ref{fig:universality} is largest at $K_{12}=K$.}

\aumk{For $K_{12}\neq K$ the two modes are separated. With} $D\to D+\Delta D\,e_1e_1^\top$\aumk{, to} first order in $\Delta D$, $G(s)=G^{0}(s)-\Delta D\,s\,\hmk{c_\ell^\top Z_0^{-1}(s)\,e_1e_1^\top Z_0^{-1}(s)\,b_r}$ with \hmk{$Z_0=Ms^2+Ds+L$}. Expanding both resolvents in the baseline modes and keeping the term resonant at \smk{\hmk{$s_+$}} gives the residue of the \admk{in-phase resonance on branch 1--2},
\begin{equation}
\smk{\hmk{R_+=\frac{\Delta D}{2m}\,(v_+^\top b_r)(v_+^\top e_1)\sum_{k\neq +}\frac{(c_\ell^\top v_k)(v_k^\top e_1)}{\Lambda_+-\Lambda_k}}}+O((\Delta D)^2),
\label{eq:sup_minimal_residue}
\end{equation}
which is \aemk{Eq.~\ref{eq:sup_residue_change} (section S4.1) written for the matrices $M$, $D$ and $L$ instead of $A$ and evaluated on the three modes}\smk{\acmk{. The} damping tilts the in-phase mode toward the antiphase mode, the only mode that carries flow on branch 1--2}, whereas the pole moves by \smk{\hmk{$\Delta s_+=-\Delta D\,(v_+^\top e_1)^2/(2m)$}} with no spacing denominator. \aumk{At the crossing $K_{12}=K$ this formula does not apply, and Eq.~\ref{eq:sup_degenerate} replaces it.} For $K_{12}=2K$ \smk{only the antiphase term remains, with \hmk{$\Lambda_+-\Lambda_-=-2K$}, so \hmk{$R_+=\Delta D/(6m)$}} and the peak of the channel at \smk{$\hmk{\omega_+}$ is \hmk{$|R_+|/|\mathrm{Re}\,s_+|$}$=2\Delta D/(6d+\Delta D)$}, linear in $\Delta D$ for $\Delta D\ll d$. At $\Delta D=d$\aomk{, with \aumk{$m=K=1$ and $d=0.04$}, the closed form agrees with the exact peak to within $1\%$, and the shift of the pole agrees with the prediction}. \smk{The antiphase resonance opens as well, through the input-side term, with \hmk{$R_-=-\Delta D/(10m)$} and a peak $2\Delta D/(10d+5\Delta D)$ at $\hmk{\omega_-}$.} 

\aumk{For large $\Delta D$ the damping holds unit 1 nearly fixed, and in Fig.~\ref{fig:minimal}C the peak of the channel rises toward $12.5$, or $87\%$ of $K/(\omega_0 d)$, at $\omega=\sqrt{K/m}$, the response of the network with unit 1 held fixed}, which is the three-unit form of the ceiling of Eq.~\ref{eq:ceiling}. With $\Delta D$ at unit 3\emk{,} \smk{\cmk{units 1 and 2 remain interchangeable} and} the channel stays at zero to all orders.\atmk{ The same holds when equal damping is added to units 1 and 2. Lowering the inertia of unit 1 or the coupling $K_{13}$ by $10\%$ instead opens the channel, as does any other change that makes units 1 and 2 differ.} \smk{The ten-node network has the same structure\acmk{. The} latent resonance at \armk{$3.37$}~Hz moves nodes \armk{4 and 10} \acmk{together}, the term that \admk{opens it on this channel} comes from the \armk{$3.42$}~Hz resonance, which moves them against each other, and \armk{the damping opens the channel by an order of magnitude} (section \umk{S4.1}).}\mkoff{}

\pon{}\paragraph{Parity under the exchange \cmk{of units 1 and 2}.} Let \hmk{$Q$} be the permutation that exchanges units 1 and 2, and let $D^{\pm}=(D_1\pm D_2)/2$ be the mean and the half-difference of their dampings. Because \hmk{\armk{$Q^\top MQ=M$, }$Q^\top LQ=L$, $Qb_r=b_r$ and $Q^\top c_\ell=-c_\ell$}, the exchange maps the network onto the one with $D^{-}$ reversed, $Z(D^{+},-D^{-};s)=\hmk{Q^\top} Z(D^{+},D^{-};s)\,\hmk{Q}$. The two networks share every eigenvalue, and $G(D^{+},-D^{-};s)=\hmk{c_\ell^\top} \hmk{Q^\top} Z^{-1}(D^{+},D^{-};s)\,\hmk{Q}\,\hmk{b_r}=-G(D^{+},D^{-};s)$. \xmk{For the three units, with $s=j\omega$,}
\begin{equation}
  \begin{aligned}
  \cmk{s_k}(D^{+},-D^{-})&=\cmk{s_k}(D^{+},D^{-}),\\
  G_{r,\ell}(D^{+},-D^{-};s)&=-G_{r,\ell}(D^{+},D^{-};s).
  \end{aligned}
  \label{eq:parity}
\end{equation}
Expanding Eq.~\ref{eq:parity} in $D^{-}$ at fixed $D^{+}$\armk{ away from its poles}, as in Eq.~\ref{eq:mirror},
\begin{equation}
G(s)=h(s)\,D^{-}+O\big((D^{-})^{3}\big),
\label{eq:sup_parity}
\end{equation}
with \hmk{$h(s)=-2s\,(c_\ell^\top \tilde Z^{-1}(s)\,e_1)(e_1^\top \tilde Z^{-1}(s)\,b_r)$}, where \aumk{$\tilde Z$ is the dynamic stiffness} of the balanced network, $D_1=D_2=D^{+}$. With several \cmk{interchangeable} pairs the same argument gives $G(s)=\sum_i h_i(s)\,D_i^{-}+O(|D^{-}|^{3})$, with $h_i$ obtained by replacing $e_1$ by $e_i$. \nmk{In the ten-node network with the inertia of nodes \armk{4 and 10} and their couplings to node \armk{1} made equal, $h\,D^{-}$ at \armk{$3.398$}~Hz reproduces the node-\armk{5} to \armk{$\ell_{4,10}$} channel to within \armk{$10^{-4}$} of its value\acmk{. The} actual channel, which also carries the small mismatch of inertia and coupling, is \armk{$66\%$} larger.} \xmk{At fixed $D^{+}$, doubling $D^{-}$ doubles the channel}\aumk{. When the resonances of the balanced network are simple, the eigenvalues, which the exchange leaves unchanged, are even in $D^{-}$ and shift at second order, so that doubling $D^{-}$ quadruples their shift\aymk{.} \aymk{A}t a crossing such as $K_{12}=K$ they split at first order.} The argument uses no property of the units beyond the exchange symmetry and holds for the linearized dynamics of each model of Fig.~\ref{fig:universality}.

A baseline that is not exactly symmetric enters through the same expansion. If unit 2 carries a damping excess \hmk{$\delta\,d$, with $\delta$ its relative excess,} and \hmk{$\Delta D$} is added at unit 1, $D^{-}$ changes from \hmk{$-\delta\,d/2$} to \hmk{$(\Delta D-\delta\,d)/2$}, and the peak of the channel changes by the factor \hmk{$\rho=|\Delta D/(\delta\,d)-1|$} to leading order. The channel closes at \hmk{$\Delta D=\delta\,d$}, where the added damping \cmk{makes units 1 and 2 interchangeable again}, and opens in proportion to $\Delta D$ for \hmk{$\Delta D\gg\delta\,d$}. For \bdmk{$K_{12}=2K$ and} the parameters above with \hmk{$\delta=10^{-3}$}, the exact ratio \aomk{agrees with this expression to within $1\%$} at \hmk{$\Delta D=10\,\delta\,d$} and \hmk{$\Delta D=50\,\delta\,d$}.\poff{}

\subsection*{Section \umk{S3}. Size and location of the openings}

\nmk{\admk{The size of the opening depends on how closely the two identically positioned nodes are matched.} With the ten-node topology, the disturbance at node \armk{5}, the output \armk{$\ell_{4,10}$} and damping added at node \armk{4} ($\varepsilon=10\%$) fixed, only the spread of the parameters is varied.} \nmk{\anmk{Figure~\ref{fig:ensemble}, E to G, shows this dependence for one parameter class at a time.} \ajmk{Table~\ref{tab:spread_scan} gives the same dependence for the random ensembles.} In table~\ref{tab:ceiling_scaling} the deviation of every inertia, damping and coupling from its mean is scaled by a common factor.}

\begin{table}[!htbp]
\centering
\caption{\textbf{\aimk{Location of the large openings.}} \acmk{Ensembles of Fig.~\ref{fig:ensemble}, $300$ networks per size, damping of one node raised by $\bar D$, best node per channel. A twin branch is a branch whose two ends have the same other neighbors, and a cell is a cell of the coarsest \ahmk{externally} equitable partition of the topology with the disturbed node alone in its cell (Materials and Methods). Shares of networks carry $95\%$ binomial confidence intervals of $\pm2$ to $\pm10$ percentage points, the wider ones for the smaller subgroups.\aumk{ Channels are counted on the frequency grid of Materials and Methods (Random network ensembles)\aymk{.} \aymk{T}he $0.01$~Hz grid of Fig.~\ref{fig:ensemble}C gives $26{,}215$ channels opened by more than a factor of $2$, against $26{,}233$ here.}}}
\label{tab:twin_control}
\vspace{0.3em}
\lmk{\begin{tabular}{lcccc}
\hline
 & $N=10$ & $N=20$ & $N=40$ & $N=80$ \\
\hline
Channels opened by more than $10$ & $32$ & $252$ & $646$ & $4342$ \\
\quad on a twin branch (\%) & $87.5$ & $98.8$ & $96.3$ & $93.6$ \\
\quad inside a cell (\%) & \qmk{$100.0$} & \qmk{$100.0$} & \qmk{$100.0$} & \qmk{$100.0$} \\
\quad damping at an end of the branch (\%) & $100$ & $100$ & $99.8$ & $98.6$ \\
Channels opened by $2$ to $10$ & $347$ & $1099$ & $3756$ & $15{,}759$ \\
\quad on a twin branch (\%) & $68.0$ & $80.8$ & $78.3$ & $75.4$ \\
\quad inside a cell (\%) & \qmk{$97.1$} & \qmk{$98.3$} & \qmk{$98.6$} & \qmk{$97.1$} \\
Networks with a twin pair (\%) & $21$ & $29$ & $45$ & $67$ \\
\quad of which with a channel opened by more than $10$ (\%) & $9.5$ & $28.4$ & $31.1$ & $65.3$ \\
Networks without a twin pair (\%) & $79$ & $71$ & $55$ & $33$ \\
\quad of which with a channel opened by more than $10$ (\%) & $1.3$ & $0.5$ & $2.4$ & $6.1$ \\
\quad largest opening ratio & $16$ & $15$ & $21$ & $22$ \\
\hline
\end{tabular}}
\end{table}%

\begin{table}[!htbp]
\centering
\caption{\textbf{Opening in the random ensembles versus parameter spread.} \acmk{Ensembles of Fig.~\ref{fig:ensemble}, $300$ realizations per size, with inertia, damping and line coupling uniform within \akmk{the listed half-width of their means (standard deviation half-width$/\sqrt{3}$)} and the damping of one node raised by the network's mean damping, maximized over nodes. \aomk{\asmk{Entries give the share of networks with a channel opened by more than a factor of $2$ (upper rows) or $10$ (lower rows).}} The column \akmk{$5\%$} is the ensemble of Fig.~\ref{fig:ensemble}.}}
\label{tab:spread_scan}
\vspace{0.3em}
\jmk{\begin{tabular}{lccccc}
\hline
Half-width & $1\%$ & $2\%$ & $5\%$ & $10\%$ & $20\%$ \\
\hline
\multicolumn{6}{l}{\asmk{Networks with a channel opened by more than $2$ (\%)}} \\
$N=10$ & \asmk{$30.0$} & \asmk{$29.3$} & \asmk{$23.3$} & \asmk{$13.3$} & \asmk{$6.0$} \\
$N=20$ & \asmk{$36.0$} & \asmk{$35.3$} & \asmk{$31.7$} & \asmk{$24.0$} & \asmk{$18.7$} \\
$N=40$ & \asmk{$54.7$} & \asmk{$54.7$} & \asmk{$52.3$} & \asmk{$47.7$} & \asmk{$37.0$} \\
$N=80$ & \asmk{$83.3$} & \asmk{$83.0$} & \asmk{$82.3$} & \asmk{$77.7$} & \asmk{$74.0$} \\
\multicolumn{6}{l}{Networks with a channel opened by more than $10$ (\%)} \\
$N=10$ & $19.7$ & $9.7$ & $3.0$ & $1.0$ & $0.3$ \\
$N=20$ & $28.0$ & $19.0$ & $8.7$ & $3.0$ & $1.7$ \\
$N=40$ & $47.0$ & $38.0$ & $15.3$ & $5.0$ & $0.7$ \\
$N=80$ & $70.0$ & $65.3$ & $46.0$ & $18.7$ & $8.0$ \\
\hline
\end{tabular}}
\end{table}%

\begin{table}[!htbp]
\centering
\caption{\nmk{\textbf{\admk{Parameter spread, opening and ceiling.}} The \armk{$\ell_{4,10}$} channel (disturbance at node \armk{5}, damping at node \armk{4}) with the deviation of every inertia, damping and coupling from its mean scaled by a common factor. Peaks are in-band maxima over \armk{$0.2$}--$10$~Hz, $\rho$ is the opening ratio for $\varepsilon=10\%$, and the ceiling peak is the limit \armk{$\Delta D_4\to\infty$} of Eq.~\ref{eq:ceiling}.}}
\label{tab:ceiling_scaling}
\begin{tabular}{ccccc}
\hline
Parameter spread & Baseline peak & Opening ratio $\rho$ & Ceiling peak & Ceiling ratio \hmk{\armk{$\chi_4$}} \\
\hline
$\pm20\%$ & \armk{$1.05\times10^{-1}$} & \armk{$3.45$} & \armk{$0.988$} & \armk{$9.4$} \\
$\pm10\%$ & \armk{$5.94\times10^{-2}$} & \armk{$5.61$} & \armk{$0.862$} & \armk{$14.5$} \\
$\pm5\%$ (baseline) & \armk{$2.96\times10^{-2}$} & \armk{$10.0$} & \armk{$0.788$} & \armk{$26.6$} \\
$\pm2.5\%$ & \armk{$1.46\times10^{-2}$} & \armk{$19.0$} & \armk{$0.751$} & \armk{$51.7$} \\
$\pm1.25\%$ & \armk{$7.18\times10^{-3}$} & \armk{$37.0$} & \armk{$0.733$} & \armk{$102$} \\
$\pm0.5\%$ & \armk{$2.85\times10^{-3}$} & \armk{$91.0$} & \armk{$0.723$} & \armk{$254$} \\
$\pm0.05\%$ & \armk{$2.83\times10^{-4}$} & \armk{$901$} & \armk{$0.716$} & \armk{$2528$} \\
\hline
\end{tabular}
\end{table}

\nmk{The opening ratio is \armk{nearly} inversely proportional to the spread\aimk{. The} product of $\rho$ and the spread in percent stays between \armk{$45$ and $50$ for spreads of $\pm5\%$ and below and reaches $69$ at $\pm20\%$} (table~\ref{tab:ceiling_scaling}), because halving the spread halves the baseline peak while the \emk{opened peak} stays between \emk{\armk{$0.26$ and $0.36$}}. In the realizations of Fig.~\ref{fig:ensemble}, E to G, the opening is set by the relative mismatch $\delta$ of the two \admk{identically positioned} nodes, with \armk{$\rho\approx0.056/\delta$} for a mismatch of their inertia or of their couplings to node \armk{1} and \armk{$\rho\approx0.64/\delta$} for a mismatch of their dampings\acmk{. The} same mismatch in \ymk{the damping \emk{\acmk{therefore} raises the baseline peak only about \armk{$9\%$} as much}}. \ammk{Over these realizations the baseline peak grows in proportion to the spread, whereas the median opened peak stays within \armk{$3\%$} of its value at identical parameters for a spread of up to \armk{$2\%$} in any of the three parameters}\atmk{ and rises by at most about a third up to $15\%$}\ammk{.} Below a spread of about \armk{$0.5\%$} in the coupling or the inertia the median opening exceeds a factor of $10$, at $5\%$ it is about \armk{$1.4$}, and at $10\%$ it falls to \armk{about $1$} \anmk{(Fig.~\ref{fig:ensemble}, E and F)}.}

\nmk{The same dependence sets the ceiling. With unlimited damping at node \armk{4}, the in-band peak of the \armk{$\ell_{4,10}$} channel saturates at the closed-form ceiling $|G^{0}-\beta\gamma/\alpha|$, \armk{$27$} times its baseline. Shrinking the spread leaves the ceiling peak nearly unchanged and lowers the baseline, so the ceiling ratio grows as the \admk{spread shrinks} (table~\ref{tab:ceiling_scaling}).}

\begin{figure}[!htbp]
  \centering
  \includegraphics[width=117mm]{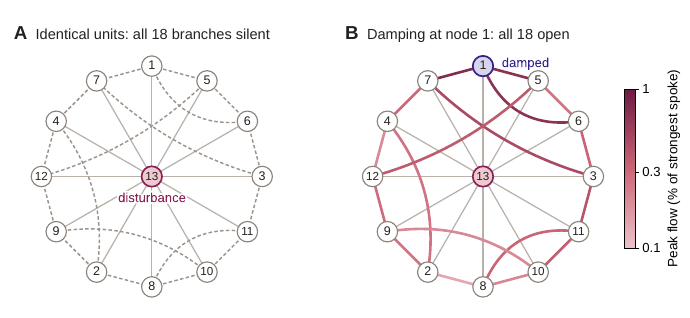}%
  \caption{\textbf{Latent channels without \ammk{graph} symmetry.} \acmk{A random $3$-regular graph on $12$ nodes whose only automorphism is the identity, with a thirteenth node \akmk{13} joined to every node. Node \akmk{13} carries the disturbance. Inertia $150$, damping $280$ and coupling $2.26\times10^{4}$ on every node and branch, band $0.2$--$10$~Hz. (\textbf{A}) Identical units. The $12$ nodes form one cell of the \ahmk{externally} equitable partition with node~13 alone in its cell, and the peak flow on each of the $18$ branches of the regular graph (dashed) is \aomk{zero to machine precision}. (\textbf{B}) Damping of $280$ added at node~1. Every branch carries flow, between $0.1\%$ and $0.8\%$ of the strongest spoke (\ammk{color}). Spokes are drawn in light gray.}}
  \label{fig:regular}
\end{figure}

\lmk{Table~\ref{tab:twin_control} reports where the \admk{large} openings of the ensembles lie. \afmk{For the \ahmk{externally} equitable partition of Materials and Methods, at identical parameters the baseline peak of every branch inside a cell is \aomk{zero to machine precision} ($N=10$ to $80$). With the $\pm5\%$ spread of the ensembles, $48$ to $64\%$ of the channels inside a cell open by more than a factor of $2$, against $0.01$ to $0.03\%$ of the others.} \gmk{An \ahmk{externally} equitable partition need not come from a symmetry of the graph\cite{Schaub2016}.} A random $3$-regular graph on $12$ nodes whose only automorphism is the identity, with a thirteenth node joined to every node as the source, has all $18$ branches of the regular graph in one cell of the \ahmk{externally} equitable partition with the source alone in its cell. At identical parameters their baseline peaks are \aomk{zero to machine precision}, and damping of $\bar D$ added at one node raises every one of them, to between $0.1\%$ and $0.8\%$ of the strongest spoke (fig.~\ref{fig:regular}). \gmk{A four-machine network with the topology of Kundur's two-area system\cite{Kundur1994}, two identical generators per area and the areas joined through a tie line, with the nominal parameters of the ensembles, behaves in the same way. For a disturbance at one generator of the first area, the two generators of the second area form a cell, and the flow between them in the network reduced to the four generators is zero \aomk{to machine precision}\acmk{. Damping} of $\bar D$ added at one of them raises it to $3.7\%$ of the strongest flow, whereas the same damping at the disturbed generator leaves it at zero.}}%

\asmk{Identical units at one plant are identically positioned in the same way. We modelled the IEEE reliability test system\cite{RTS1979,Grigg1999}, its 2019 update\cite{Barrows2020} and the synthetic 2000-bus Texas grid\cite{Birchfield2017}, with the data of MATPOWER\cite{Zimmerman2011}, each synchronous unit as its own node behind a transient reactance of $0.25$ on its rating, inertia constants typical of its fuel, a spread of $1\%$ in the inertia of units of the same type, and damping scaled to a median modal damping ratio of $3\%$. Units of the same type and rating at one bus then form a cell. With the disturbance at one unit of such a plant and the network's mean unit damping added at another, the power difference between \azmk{the damped unit and an identical unit of the same plant} opens by a median factor of $8$, $11$ and $7$ in the three grids, to a median of $15\%$, $15\%$ and $8\%$ of the strongest flow, at the modes in which the units of one plant swing against each other\azmk{, known as intra-plant modes}\cite{Kundur1994} ($2.1$ to $2.5$~Hz in the two reliability test systems). No transmission line opens by more than a factor of $2.5$, as identically positioned nodes occur in these grids only within plants.}%

\ahmk{In the ensembles of Fig.~\ref{fig:ensemble}, with the damping of one node raised by $\bar D$, the channels opened by more than a factor of $2$ carry after the opening a median of $0.5\%$ \aomk{and} a maximum of $27\%$ of the strongest baseline flow under the same disturbance, over all sizes. \apmk{No channel in any network falls below $29\%$ of its baseline peak.}}%

\subsection*{Section \umk{S4}. Mechanism of the residue change}

\subsubsection*{\umk{S4.1} \aimk{Dominance of the antiphase mode in the residue change}}
The sensitivity of a channel residue to damping follows from the first-order perturbation of the eigenvector \hmk{$\hat v_k$ of $A$ (left eigenvector $\hat w_k$)} under a rank-one damping change \hmk{$\partial A/\partial(\Delta D)$},
\begin{equation}
\hmk{\frac{\partial \hat v_k}{\partial(\Delta D)}=\sum_{q\neq k}\frac{\hat w_q^\top(\partial A/\partial(\Delta D))\,\hat v_k}{s_k-s_q}\,\hat v_q} ,
\label{eq:sup_eigvec_perturbation}
\end{equation}
\nmk{so a small spacing \hmk{$s_k-s_q$} or a large numerator lets the same dissipation rotate the eigenvector strongly and rewrite the residue \hmk{$(\hat c^\top\hat v_k)(\hat w_k^\top\hat b)$}, whereas the pole velocity \hmk{$\hat w_k^\top(\partial A/\partial(\Delta D))\hat v_k$} has neither. With $\hat w_q^\top\hat v_k=\delta_{qk}$ and $A'=\partial A/\partial(\Delta D)$, the residue $R_k=(\hat c^\top\hat v_k)(\hat w_k^\top\hat b)$ changes at first order as
\begin{equation}
\frac{\partial R_k}{\partial(\Delta D)}=\sum_{q\neq k}\frac{(\hat w_q^\top A'\hat v_k)(\hat c^\top\hat v_q)(\hat w_k^\top\hat b)+(\hat c^\top\hat v_k)(\hat w_k^\top A'\hat v_q)(\hat w_q^\top\hat b)}{s_k-s_q} ,
\label{eq:sup_residue_change}
\end{equation}
where the first term carries the flow $\hat c^\top\hat v_q$ of resonance $q$ on the monitored branch. %
 In the random ten-node network (disturbance at node \armk{5}, branch \armk{$\ell_{4,10}$}, damping at node \armk{4}), the first-order change of the residue of the \armk{$3.37$}~Hz resonance with \armk{$\Delta D_4$} is expanded term by term in Eq.~\ref{eq:sup_residue_change}. The \armk{$3.42$}~Hz resonance and its conjugate give \armk{$99.8\%$} of it\armk{. This resonance is also the nearest, at about one-seventh of the distance to the next}. \aimk{In this expansion of the residue change, the} numerator of the \armk{$3.42$}~Hz term is about \armk{$40$} times larger than that of any other term\armk{, and the resonance} carries about \armk{$26$} times more flow on \armk{$\ell_{4,10}$} than any other resonance\acmk{. It} moves nodes \armk{4 and 10} against each other and keeps \aomk{almost all} of its mode vector on the pair, whereas the \armk{$3.37$}~Hz resonance moves the two nodes \acmk{together}, with \aomk{nearly equal components}.}

\nmk{Three checks show that the decomposition holds at the actual dose. First, \aomk{a reduced model that keeps only the \armk{$3.37$ and $3.42$}~Hz modes and their conjugates \armk{gives}} a residue amplification of \armk{$7.6$} at $\varepsilon=10\%$, against \armk{$7.7$} for the full model\armk{. The first-order extrapolation overshoots, to $16$, because at this dose the $3.42$~Hz pole moves by about twice the spacing of the two poles}\bcmk{, so that at this dose the opening is described by the reduced model, and Eq.~\ref{eq:sup_residue_change} gives its direction}. Second, without the \armk{$3.42$}~Hz term the first-order extrapolation gives an amplification of only \armk{$1.00$}. Third, the coupling between the two modes is proportional to the product of their components at the damped node, and splitting the same damping equally between nodes \armk{4 and 10} reduces it by a factor of \armk{$10$}, in agreement with the absence of opening under symmetric placement \kmk{(Eq.~\ref{eq:sup_parity})}.} Equation~\ref{eq:sup_eigvec_perturbation} is a first-order result and fails at exact degeneracy, where degenerate perturbation theory is required.

\subsubsection*{\umk{S4.2} \aimk{Input and output selectivity}}
\begin{figure}[!htbp]
  \centering
  \includegraphics[width=516.1bp,trim=0bp 1.5bp 0bp 2.4bp,clip]{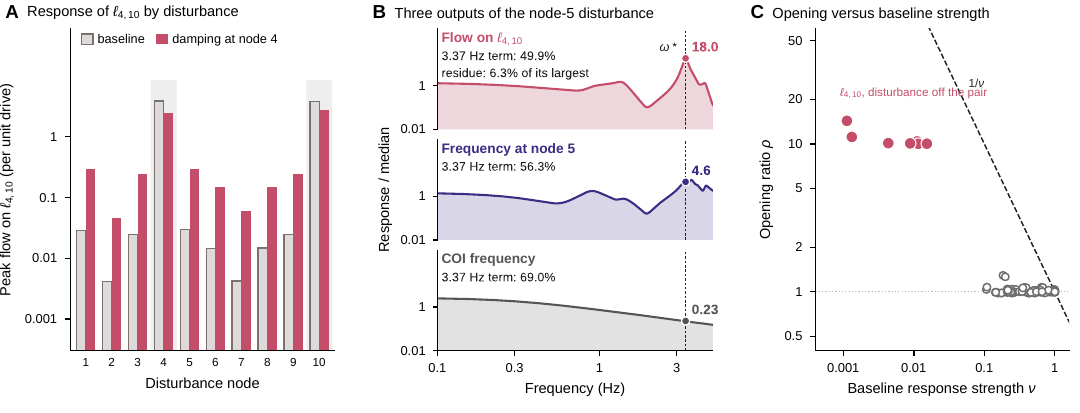}%
  \caption{\textbf{Input and output selectivity of a latent resonance.} \acmk{(\textbf{A}) Peak of the \armk{$\ell_{4,10}$} flow over \armk{$0.2$}--$10$~Hz for a unit sinusoidal disturbance at each node, at baseline (gray bars) and with $\varepsilon=10\%$ added at node \armk{4} (red bars). The gray \armk{bands mark} disturbances on the pair. (\textbf{B}) Baseline responses of three outputs to the node-\armk{5} disturbance, each normalized by its median over $0.1$--$5$~Hz: the flow on \armk{$\ell_{4,10}$} (red), the frequency at node \armk{5} (blue) and the center-of-inertia frequency (gray). The dashed line marks the baseline $\omega^\star/2\pi=\armk{3.398}$~Hz, and dots mark the three responses there, with their values. Percentages give the share of the \armk{$3.37$}~Hz term among the oscillatory modal terms of each response at $\omega^\star$. The residue is that of the \armk{$3.37$}~Hz resonance on the flow channel relative to its largest residue \ahmk{(section S4.2)}. (\textbf{C}) Opening ratio $\rho$ versus baseline response strength $\nu$ for all \armk{$110$} \akmk{disturbance}--branch channels. \ammk{Circles mark the largest $\rho$ of each channel over the ten damping locations,} and red dots $\rho>2$, labeled by their branch\armk{ and disturbance}. The dotted line marks $\rho=1$, and the dashed line $1/\nu$, at which a channel would match the strongest baseline response.}}
  \label{fig:hidden-response-channels}
\end{figure}

\nmk{Figure~\ref{fig:hidden-response-channels} shows the selectivity of the latent resonance. Damping at node \armk{4} opens the \armk{$\ell_{4,10}$} channel for every disturbance off the pair, by a median factor of \armk{$10$}, and \dmk{lowers it by about a third} for disturbances on the pair (A). At \hmk{the baseline} $\omega^\star$ the \armk{$3.37$}~Hz resonance dominates the frequency at node \armk{5} and at the center of inertia, whereas its residue on the \armk{$\ell_{4,10}$} flow is \armk{$6.3\%$} of its largest residue (B). Over all \armk{$110$} \akmk{disturbance}--branch channels, the channels that open most are the weakest at baseline (C). \apmk{For the disturbance at node \armk{5}, the opened channel carries \armk{$15\%$} of the flow on the strongest branch, \armk{$\ell_{3,9}$}.}}

\mkon{}For each of the \fon{}$26{,}215$\foff{} ensemble channels opened by more than a factor of $2$ (Fig.~\ref{fig:ensemble}C), the resonance nearest the post-intervention worst-case frequency is matched to its baseline pole $\cmk{s_k}$\acmk{. Its} baseline residue share is \ajmk{split into} the output factor \hmk{$|c_\ell^\top v_k|$, relative to its largest value over the branches of the network,} and the input factor \hmk{$|w_k^\top b_r|$, relative to its largest value over its nodes}. The output factor is the smaller of the two on \fon{}$97\%$\foff{} of the channels, with a median of $2.2\times10^{-3}$ against $0.38$ for the input \aomk{factor}. \nmk{The random ten-node network follows the same pattern, with \emk{\armk{$6.3\times10^{-2}$}} at the output and $1$ at the input, \ahmk{the residue share of \armk{$6.3\%$} of fig.~\ref{fig:hidden-response-channels}B}.}\mkoff{}%

\subsubsection*{\wmk{\umk{S4.3} \aimk{Stored energy and transient growth}}}

Non-normality can transiently amplify small deviations in a linearly stable network, and this amplification, measured by the reactivity, cannot be explained by the spectrum\cite{Trefethen1993,Neubert1997,Asllani2018,Duan2022}. In the energy norm the reactivity of the network is zero before and after any damping increment, because the stored energy $E=\tfrac12\dot\theta^\top M\dot\theta+\tfrac12\theta^\top L\theta$ obeys $\dot E=-\dot\theta^\top D\dot\theta\le 0$, so transient growth does not produce the amplification. \aimk{In the ensembles of Fig.~\ref{fig:ensemble}, with the damping of one node raised by $\bar D$, the energy stored in the network under white-noise power injection at the disturbed node, $\tfrac12\langle\dot\theta^\top M\dot\theta+\theta^\top L\theta\rangle$ obtained from a Lyapunov equation, falls for all $2.55\times10^{6}$ combinations of network, disturbed node and damped node. Over all $45{,}000$ single-node increments, every node of every network, \awmk{the increment improves the spectrum slightly in nearly all cases, and the smallest damping ratio rises in $99.7\%$ of them.}%
}

\subsection*{Section \umk{S5}. Conventional screening indices}
\begin{figure}[!tbp]
  \centering
  \includegraphics[width=396.85bp]{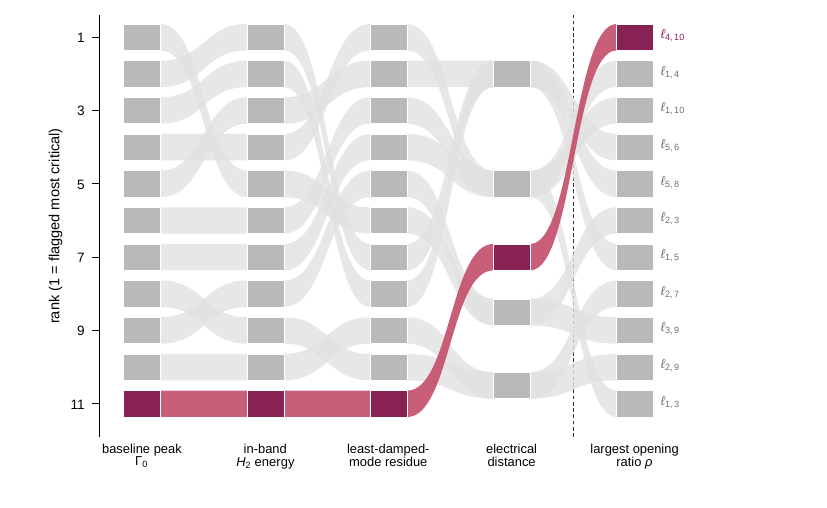}%
  \caption{\textbf{Channel rankings by conventional screening indices.} \acmk{Ranks of the \armk{$11$} branch channels of the ten-node network under the disturbance at node \armk{5} by the baseline peak gain, the in-band $H_2$ energy\ammk{\cite{Tyloo2019}}, the residue of the least-damped in-band mode\ammk{\cite{Kundur1994,Pagola1989}} and the electrical distance\ammk{\cite{Tyloo2019}} ($1$ = flagged most critical). The rightmost column gives the ranking by \ammk{the largest opening ratio $\rho$} over single-node interventions of $\varepsilon=10\%$. Ribbons join the ranks of the same channel, with the \armk{$\ell_{4,10}$} channel in red. Tied values share their average rank.}}
  \label{fig:competing-indices}
\end{figure}
\nmk{For the disturbance at node \armk{5} and all \armk{$11$} branch outputs of the ten-node network, four conventional indices are computed per channel: the baseline peak gain $\Gamma_0$, the in-band $H_2$ energy \hmk{$\int_{\mathcal{W}}|G^{0}_{r,\ell}|^2\,\mathrm{d}\omega$}, which is the channel version of performance-based key-player ranking\cite{Tyloo2019}, the channel residue \armk{$|c_\ell^\top v_k||w_k^\top b_5|$} of the least-damped in-band mode, the one with the largest real part\cite{Kundur1994,Pagola1989}%
, and the electrical distance\ammk{\cite{Tyloo2019}}\aimk{, the effective resistance between the disturbed node and the nearer end of the branch, computed from the pseudoinverse of the Laplacian}. \ammk{For each channel, the largest opening ratio $\rho$ is taken} over single-node interventions of $\varepsilon=10\%$.}

\nmk{\armk{Three of the} four indices rank the \armk{$\ell_{4,10}$} channel last among the \armk{eleven, and the electrical distance ranks it seventh} (fig.~\ref{fig:competing-indices}). The participation factor of the \armk{$3.37$}~Hz resonance is largest at node \armk{5 ($1.00$) and node 1 ($0.82$), and $0.22$ and $0.21$ at nodes 4 and 10}. It ranks node \armk{5} first, although damping there lowers the channel (\armk{$\rho=0.77$}), and places nodes \armk{4 and 10}, the only two that open it by more than $30\%$, \armk{seventh and eighth}. The sensitivity of the \armk{$3.37$}~Hz pole to the damping of each node gives the same ranking. \ammk{With $\varepsilon=10\%$ at each node, the damping ratio of this resonance rises by \armk{$24\%$ for damping at node 5, $20\%$ at node 1 and $0.7\%$ and $1.8\%$ at nodes 4 and 10}.}}

\subsection*{Section \umk{S6}. Pan-European census}
\begin{figure}[!tbp]
  \centering
  \includegraphics[width=476.0bp]{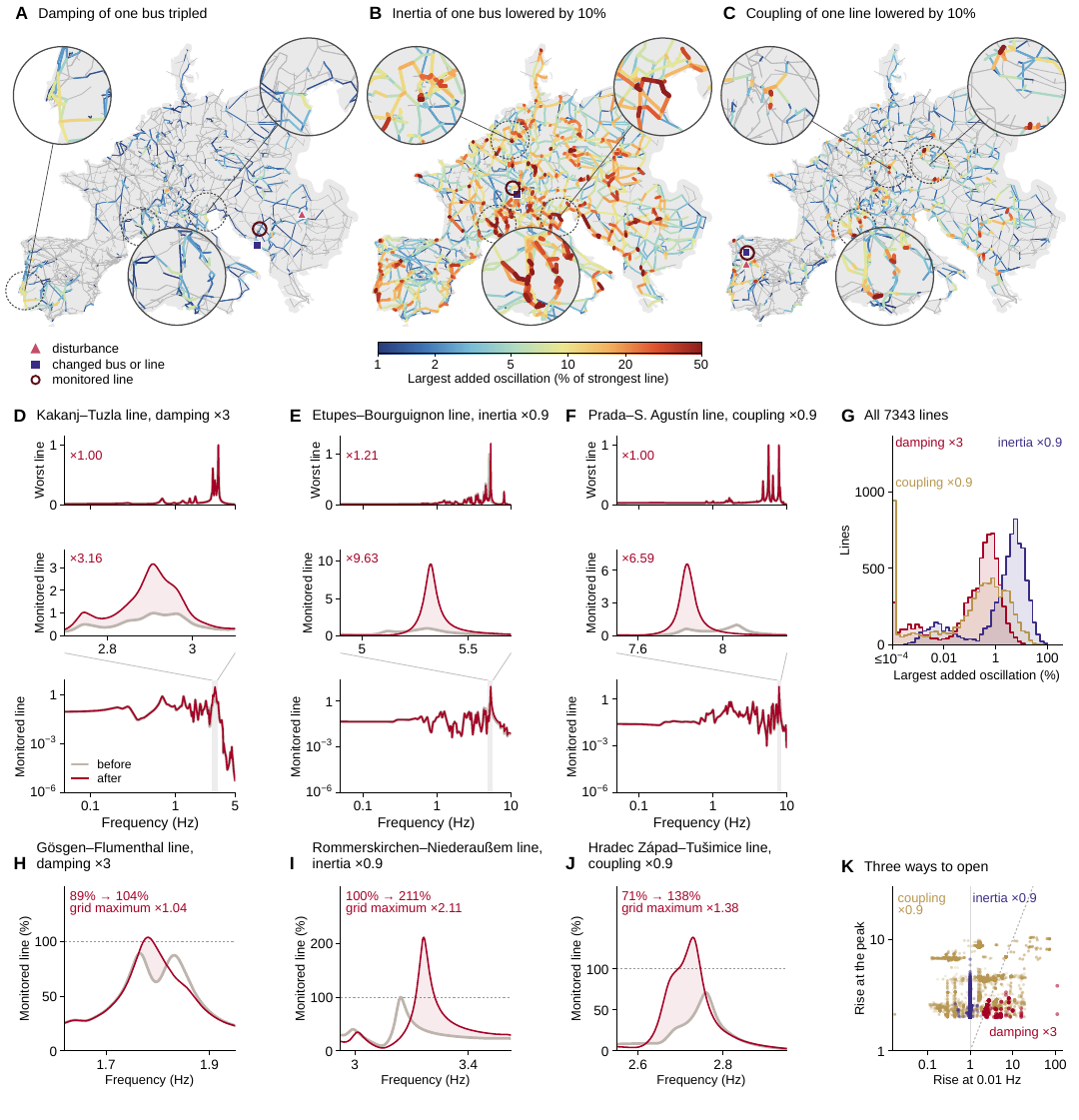}%
  \caption{\textbf{Response of the pan-European grid to \asmk{local damping, local inertia and local line coupling}.} \acmk{PanTaGruEl model with $3809$ buses and $7343$ lines. Each of the $468$ buses with inertia serves as disturbance source and as changed bus, all lines are monitored over $0.05$--$10$~Hz, and all combinations are counted. \asmk{For the line coupling, each of the $7343$ lines is changed in turn and only the lines that share an end bus with it are monitored, without the changed line and its parallel circuits, so (C) and the coupling histogram of (G) are lower bounds (section \bcmk{S12}).} (\textbf{A} \asmk{to \textbf{C}}) Largest added oscillation of each line over all sources and changed buses \asmk{or lines}, relative to the strongest line of the same source, when the damping of one bus is tripled (A)\asmk{,} its inertia \asmk{lowered by $10\%$} (B)\asmk{, or the coupling of one line lowered by $10\%$ (C)}. \asmk{Lines below $1\%$ are shown in gray, and colors saturate at $50\%$.} Circles magnify the three regions with the largest added oscillation in each map, and markers show the cases of \asmk{(D) to (F)\aymk{.} \aymk{I}n (B) the disturbance and the changed bus coincide}. (\textbf{\asmk{D}}) Disturbance at Paroseni (Romania), damping tripled at \v{C}apljina (Bosnia and Herzegovina), monitored line Kakanj--Tuzla (Bosnia and Herzegovina). The top panel shows the worst line of the grid, the middle panel the monitored line in the gray band below, and the bottom panel the monitored line over $0.05$--$5$~Hz. Gray and red show the responses before and after the change. (\textbf{\asmk{E}}) As (\asmk{D}), \asmk{for a disturbance at M\"{u}hleberg (Switzerland), inertia lowered by $10\%$ at the same bus and the monitored line Etupes--Bourguignon (France), whose peak lies at $5.3$~Hz\aymk{.} \aymk{T}he bottom panel extends to $10$~Hz.} \asmk{(\textbf{F}) As (D), for a disturbance at Baixo Sabor (Portugal), the coupling of the P.~Bibey--Prada line (Spain) lowered by $10\%$ and the monitored line Prada--S.~Agust\'{\i}n (Spain), whose peak lies at $7.8$~Hz\aymk{.} \aymk{T}he bottom panel extends to $10$~Hz.}  (\textbf{\asmk{G}}) Histogram of the largest added oscillation of each line for damping tripled\asmk{,} inertia \asmk{lowered by $10\%$} \asmk{and coupling lowered by $10\%$}. The leftmost bin collects values of $10^{-4}\%$ or less. \asmk{(\textbf{H} to \textbf{J}) The combination with the largest added oscillation for each change, monitored line in percent of the strongest baseline line (dashed): disturbance at G\"{o}sgen (Switzerland), damping tripled at K\"{u}hmoos (Germany), line G\"{o}sgen--Flumenthal (H); disturbance at Goldenberg (Germany), inertia \asmk{lowered by $10\%$} at Niederau\ss{}em (Germany), line Rommerskirchen--Niederau\ss{}em, which ends at the changed bus (I); disturbance at Tisov\'{a} (Czech Republic), coupling of the Hradec Z\'{a}pad--V\'{y}\v{s}kov line lowered by $10\%$, line Hradec Z\'{a}pad--Tu\v{s}imice (J). Each line responds strongly before the change and \bdmk{is} the strongest of the grid after it.} (\textbf{\asmk{K}}) Rise at the peak versus rise at $0.01$~Hz, each after over before, for the resonant channels \asmk{below $2$~Hz opened by more than a factor of $2$}\bdmk{ by tripled damping (red), inertia lowered by $10\%$ (indigo, a random sample of $6000$) and coupling lowered by $10\%$ (gold)}\bcmk{, without the $1\%$ screening of the census}. The dashed line marks equal gains.}}
  \label{fig:real-grid}
\end{figure}
\ron{}\bcmk{On the pan-European grid, local damping leaves the spectral margin nearly unchanged and opens channels by up to a factor of about $3$, at resonances that are weak on these channels at baseline.}

The PanTaGruEl model\kmk{\cite{Pagnier2019,Tyloo2019}} contains $3809$ buses, $7343$ lines at $200$~kV and above, and $618$ generating units (fig.~\ref{fig:real-grid}A). Following the example scripts distributed with the dataset, generation is dispatched by a DC optimal power flow, and the AC real power flow of the lossless network \hmk{(line susceptances equal to the inverse reactances, unit voltage magnitudes)} gives the operating point, with a largest angle difference of $22.7^\circ$. Inertia is taken from the field \texttt{gen\_inertia}, and nodal damping is the sum of \texttt{gen\_prim\_ctrl} and \texttt{load\_freq\_coef}.\bcmk{ The coefficient \texttt{load\_freq\_coef} is positive at every bus, so $D\succ0$.} \aemk{At this operating point, $468$ of the $3809$ buses carry inertia. Between the 5th and 95th percentiles, the inertia of these buses varies by a factor of $38$, the nodal damping by $105$ and the linearized line coupling by $26$.}%

All $468$ buses with inertia serve as disturbance sources and as candidate damping locations\kmk{, where the damping is raised by $\kappa=2$ times its own value, that is, tripled}. The disturbance is a sinusoidal power injection at the source bus. Because the analysis is linear, every reported quantity is a ratio and does not depend on its amplitude. %
\admk{\akmk{This increment,} \hmk{$\Delta D_i=\kappa D_i$}\akmk{,} amounts\aemk{, at the median over the $468$ generator buses,} to $0.145\%$ of the network total. At the median bus \hmk{$D_i$} corresponds to $21$~MW/Hz, so $\kappa=2$ adds $43$~MW/Hz, \hmk{the frequency response, rating divided by droop and nominal frequency, of a unit of about $107$~MW with $5\%$ droop at $50$~Hz}\cite{Kundur1994}. Responses are evaluated at $800$ logarithmically spaced frequencies between $0.05$ and $10$~Hz. For every disturbance source and monitored line, the opening ratio is taken at the damping location that maximizes it, and $\nu$ is normalized by the largest baseline peak over all lines under the same source.}

A channel counts as resonant when its \emk{peak $\Gamma$ over $0.05$--$10$~Hz} exceeds twice its \azmk{response at $0.01$~Hz, $\Pi$, computed separately from the frequency grid}, both before (subscript $0$) and after \hmk{(prime)} the intervention, with \hmk{$\Pi'$} evaluated at the damping location that produces \hmk{$\Gamma'$}. \aumk{\azmk{Combinations} whose opened peak stays below $1\%$ of the strongest baseline line under the same disturbance are left out.} Combinations with the damping at an end of the monitored line are kept. The common factor is the median opening ratio of the remaining combinations, and a deviation is an opening ratio more than $20\%$ above it. \aimk{The most strongly responding channels are the fifth of the combinations with the largest $\nu$.} Statistics per corridor use its combination with the largest opening ratio. Parallel circuits between the same end buses are counted as one corridor. The worst response over all lines is evaluated with the damping at the bus that opens the monitored line. The spectral margin is the largest real part among the nonzero eigenvalues of the descriptor pencil, obtained by shift-invert iteration at $41$ shifts between $0$ and $50$~Hz before and after each of the $468$ single-bus increments. \admk{Identically positioned buses are taken from} \qmk{the coarsest \ahmk{externally} equitable partition, computed with the refinement of Materials and Methods,} of the topology \hmk{with parallel circuits kept as separate edges, buses with and without inertia distinguished and} all \hmk{other} parameters set equal. The dose sweep \jmk{in the last row of table~\ref{tab:census}} uses $100$ disturbance sources drawn at random without replacement and the common factor of each dose. The representative case is recomputed on a frequency grid of $0.0005$~Hz around its peak.\roff{}%

Table~\ref{tab:census} summarizes the census of fig.~\ref{fig:real-grid}. \emk{Poles in $0.2$--$3$~Hz} are obtained by sparse shift-invert iteration and their \chgon{}mode vectors by inverse iteration. For each deviating corridor whose opened peak lies in $0.2$--$3$~Hz, \aimk{the residue share of the pole $s_k$ in this band nearest to the peak is its residue on the channel relative to its largest residue over all lines and disturbance sources. It equals the product of $|c_\ell^\top v_k|$ and $|w_k^\top b_r|$, each relative to its largest value over all lines or all disturbance sources}\chgoff{}.

\textbf{\aimk{Openings across the grid.}} \aimk{\bdmk{Figure}~\ref{fig:real-grid}A shows the added oscillation of every line without screening. \asmk{The $13$ lines raised by more than a tenth of the strongest line already respond strongly at baseline: they start at a median of $36\%$ of it and reach $48\%$, opening by at most a factor of $1.47$.} \aumk{Of the $6.38\times10^{5}$ source--line combinations that are resonant before and after the change and reach at least $1\%$ of the strongest line, three open by more than a factor of $2$, the largest by $3.12$ (table~\ref{tab:census}). The channels that respond most strongly at baseline open by at most $1.55$ under the same procedure. On the corridors that deviate from the typical change, the resonance at the opened peak is weaker at baseline: its residue has a median of $7.3\times10^{-3}$ of its largest residue, against $2.7\times10^{-2}$ on the most strongly responding channels (table~\ref{tab:census}).}} \aimk{\bdmk{The openings on this grid are smaller than in the ten-node network. The grid's parameters are heterogeneous, and almost no two buses are identically positioned (table~\ref{tab:census}),} whereas the \armk{$\ell_{4,10}$} channel of the ten-node network opens far more once that network's parameters are nearly uniform (Fig.~\ref{fig:ensemble}, E and F).}

\textbf{\aimk{Global indicators.}} Under a disturbance in Romania, damping added at a bus in Bosnia and Herzegovina raises the resonant peak on the Kakanj--Tuzla corridor, the largest opening of the census, by a factor of $3.16$ \dmk{\emk{when} recomputed on a finer frequency grid around its peak ($3.12$ on the census grid\emk{\acmk{,} fig.~\ref{fig:real-grid}\asmk{D}})}. The spectral margin changes by less than $1\%$ under every single-bus increment\dmk{\acmk{. As} a property of the whole spectrum, it does not register a change confined to one corridor}. \asmk{A single increment can even raise the largest line oscillation of the grid while the oscillation summed over all lines falls. For $100$ sampled sources and each of the $468$ buses, the largest line peak rises by more than $0.1\%$ in $261$ of the $46{,}800$ combinations, by up to $6.1\%$\atmk{, on the Gy\H{o}r--Gab\v{c}\'{\i}kovo line under a disturbance at Mochovce (Slovakia) with the damping tripled at Chvaletice (Czech Republic)}, and in $88\%$ of these the sum of squared line peaks falls. Under a disturbance at G\"{o}sgen (Switzerland), tripling the damping at K\"{u}hmoos (Germany) lowers the strongest line to $54\%$ of its baseline peak and raises the G\"{o}sgen--Flumenthal line from $89\%$ to $104\%$ of it (fig.~\ref{fig:real-grid}H).} \xmk{The model lumps primary control and the frequency sensitivity of loads into a constant nodal \azmk{damping.}}

\begin{table}[!htbp]
\centering
\caption{\textbf{Pan-European census at $\kappa=2$.} \acmk{Criteria as described in this section. Percentages refer to the resonant combinations.}}
\label{tab:census}
\vspace{0.3em}
\begin{tabular}{lc}
\hline
Source--line combinations & $3.44\times10^{6}$ \\
Resonant & $3.15\times10^{6}$ ($91.6\%$) \\
\aumk{Opened peak below $1\%$ of the strongest line} & \aumk{$79.7\%$} \\
\aumk{Remaining} & \aumk{$6.38\times10^{5}$} \\
Opening ratio, median / $99$th percentile & \aumk{$1.008$ / $1.207$} \\
$\rho>1.5$ / $\rho>2$ & \aumk{$221$} / $3$ \\
\qmk{\ahmk{Externally equitable}-partition classes} (nontrivial; buses in them) & \qmk{$3748$} ($58$; \qmk{$119$}) \\
\emk{Poles in} $0.2$--$3$~Hz (median damping ratio) & $181$ ($0.0125$) \\
Spectral margin & $4.21$~Hz, $-0.101$~s$^{-1}$ \\
Margin shift toward / away from the axis & $5\times10^{-6}$ / $6.4\times10^{-4}$~s$^{-1}$ \\
\chgon{}Deviating corridors, $468$ sources & \aumk{$1474$} \\
\chgon{}Largest opening ratio & \chgon{}$3.12$ \\
\emk{Deviating corridors with the opened peak below $2$~Hz (largest $\rho$)} & \aumk{$961$} ($1.99$) \\
\chgon{}Residue share at the opened peak, median ($<1\%$; corridors) & \aumk{$7.3\times10^{-3}$ ($59\%$; $1346$)} \\
\aimk{Most strongly responding channels, largest $\rho$ / median residue share} & \aumk{$1.55$ / $2.7\times10^{-2}$} \\
Deviating corridors, $100$ sampled sources, $\kappa=0.5$ / $2$ / $10$ & \aumk{$3$ / $680$ / $2672$} \\
\hline
\end{tabular}
\end{table}

\subsection*{Section \umk{S7}. \kmk{Random distributions of the added damping}}
\begin{figure}[!htbp]
  \centering
  \includegraphics[width=354bp,trim=4bp 4bp 4bp 4bp,clip]{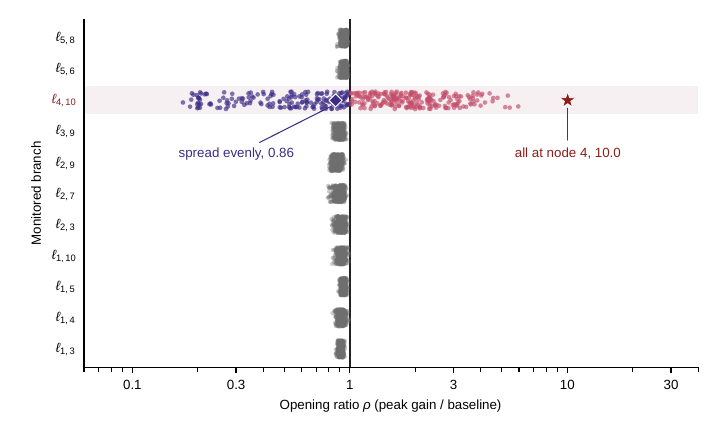}%
  \caption{\textbf{Opening under random distributions of the added damping.} \acmk{Ten-node network with the added budget $B=\varepsilon\sum_nD_n$, $\varepsilon=10\%$, distributed over the ten nodes along $400$ random Dirichlet directions (concentration parameter $1$, uniform on the simplex). The protocol is otherwise that of Fig.~\ref{fig:local-damping-fragility}. Each row is one monitored branch, and each dot is one allocation placed at its opening ratio $\rho$ on a logarithmic axis, with vertical jitter. In the \armk{$\ell_{4,10}$} row, red dots open the channel ($\rho>1$) and blue dots lower it. The diamond marks the budget spread evenly, and the star the whole budget at node \armk{4}. The black line marks $\rho=1$. The \armk{$\ell_{4,10}$} channel opens in \armk{$56\%$} of the allocations and more than doubles in \armk{$24\%$}, with $\rho$ between \armk{$0.17$ and $5.9$}\aymk{.} \aymk{T}he other \armk{ten} channels stay between \armk{$0.79$ and $0.99$}.}}
  \label{fig:randdir}
\end{figure}

\nmk{The main text places the damping at each of the ten nodes in turn. To test whether the opening depends on concentrating the budget at a single node, the total $B=\varepsilon\hmk{\sum_nD_n}$ with $\varepsilon=10\%$ is \dmk{added to the baseline and} distributed over the ten nodes along $400$ random Dirichlet directions, with the protocol otherwise unchanged (fig.~\ref{fig:randdir}). The \armk{$\ell_{4,10}$} channel opens with a median ratio of \armk{$1.18$}\aomk{,} \armk{$24\%$} of the directions exceed a factor of $2$, and the largest reaches \armk{$5.95$}. The ratios of the other \armk{ten} channels never exceed \armk{$0.99$} in any of the $400$ directions, so the selectivity does not depend on a particular placement. Spreading the \dmk{added budget evenly} over the ten nodes reduces the channel to \armk{$0.86$} times its baseline, and giving it all to node \armk{5}, the disturbed node, to \armk{$0.77$} times. The same budget lowers the channel in \armk{$44\%$} of the \aomk{directions}. The largest random direction stays below the single-node value of \armk{$10.0$} because Dirichlet directions rarely concentrate the weight on one node.}

\subsection*{Section \umk{S8}. Same-budget designs}
\anmk{For the named designs of Fig.~\ref{fig:channel_aligned_reallocation}, with the total damping held at its baseline value, the opening ratio $\rho$ of the \armk{$\ell_{4,10}$} channel under the disturbance at node \armk{5} is \armk{$0.438$} for uniform damping, \armk{$0.824$} for the footprint-based allocation, \armk{$0.278$} for the topology-based allocation and \armk{$10.7$} for the mistuned allocation, \armk{and $9.95$ with} the dampings exchanged within each pair. All designs satisfy the budget and device bounds and are linearly stable.}

\chgon{}

\subsection*{Section \umk{S9}. Robustness to the damping of the model}
All nodal damping of the model is multiplied by a common factor\dmk{, and the added damping, $\kappa=2$ times the nodal damping, scales with it. The test uses} \jmk{$100$ sampled disturbance sources}\dmk{,} $400$ logarithmically spaced frequencies between $0.05$ and $10$~Hz and the screening of \ron{}section \umk{S6}\roff{}. A factor of $2.4$ raises the median damping ratio of the $181$ poles \emk{in $0.2$--$3$~Hz} from $0.0125$ to $0.030$. \bcmk{On this grid, the combinations opened by more than a factor of $2$ rise from $3$ to $11$ and the deviating corridors from $689$ to $782$, while the largest opening ratio falls from $3.16$ to $2.83$. At the model damping, the census grid of $800$ frequencies gives $680$ deviating corridors for the same sources.}

\chgoff{}

\subsection*{\dmk{Section S10. Sustained oscillation of a bistable receiving unit}}
\begin{figure}[!htbp]
  \centering
  \includegraphics{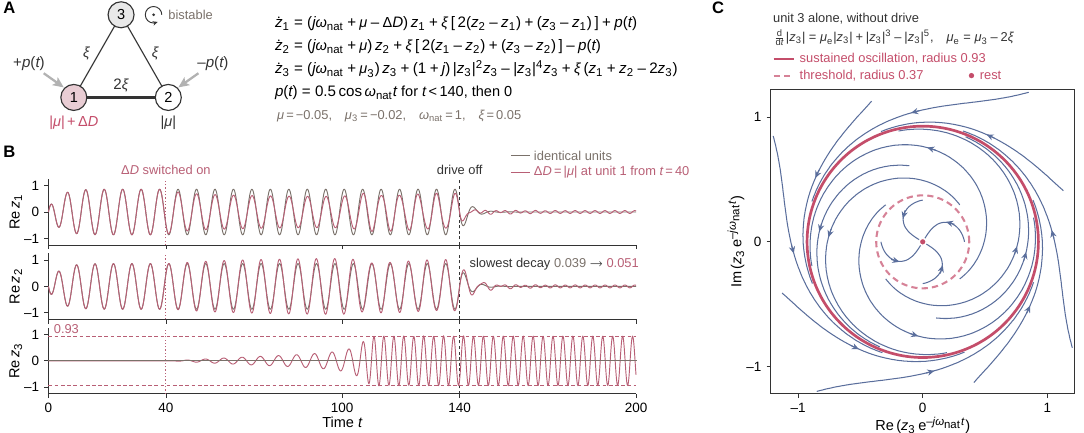}%
  \caption{\textbf{Switching of a bistable unit into sustained oscillation through a latent channel.} \acmk{(\textbf{A}) Units 1 and 2 are linear damped oscillators driven with opposite signs, and unit 3 is bistable\cite{KatoKori2024}. (\textbf{B}) Real parts of $z_1$, $z_2$ and $z_3$ for identical units (gray) and with $\Delta D=|\mu|$ added at unit 1 at $t=40$ (red, dotted line). The vertical dashed line marks the end of the drive at $t=140$, and the horizontal dashed lines the radius $0.93$ of the sustained oscillation. (\textbf{C}) Phase portrait of unit 3 alone, with growth rate $\mu_{\mathrm{e}}$, without drive, in the frame rotating at $\omega_{\mathrm{nat}}$: rest state (dot), threshold cycle of radius $0.37$ (dashed) and sustained oscillation of radius $0.93$ (solid).}}%
  \label{fig:bistable}
\end{figure}

\dmk{Units 1 and 2 of \aumk{a triangle as in} Fig.~\ref{fig:minimal} are linear damped oscillators, \hmk{$\dot z_i=(\mu_i+j\omega_{\mathrm{nat}})z_i+\xi\sum_nK_{in}(z_n-z_i)+p_i$}, as model E of table~\ref{tab:osc_models} without its cubic term. Unit 3 is the extended Stuart--Landau oscillator of ref.~\cite{KatoKori2024}, \hmk{$\dot z_3=(\mu_3+j\omega_{\mathrm{nat}})z_3+(1+j)|z_3|^2z_3-|z_3|^4z_3+\xi\sum_nK_{3n}(z_n-z_3)$}, which on its own is bistable for \hmk{$-1/4<\mu_3<0$} (fig.~\ref{fig:bistable}A). With \hmk{$\mu_1=\mu_2=\mu=-0.05$}, \hmk{$\omega_{\mathrm{nat}}=1$}, $\xi=0.05$, $K_{12}=2$, $K_{13}=K_{23}=1$ and \hmk{$\mu_3=-0.02$}, the coupling to units 1 and 2 gives unit 3 the effective growth rate \hmk{$\mu_{\mathrm{e}}=\mu_3-2\xi=-0.12$}. Unit 3 \emk{alone, with growth rate $\mu_{\mathrm{e}}$,} then has a stable rest state, a stable cycle of radius \hmk{$0.93$} and an unstable cycle of radius \hmk{$0.37$} between them (fig.~\ref{fig:bistable}C). The drive \hmk{$p_1=-p_2=0.5\cos\omega_{\mathrm{nat}}t$} acts for $t<140$. This is the channel of Fig.~\ref{fig:minimal} with input and output exchanged. With identical units 1 and 2, exchanging them reverses the sign of the drive and leaves the network unchanged, so the response of unit 3 equals its own negative and vanishes (fig.~\ref{fig:bistable}B, gray).}

\dmk{Doubling the decay rate of unit 1, \hmk{$\mu_1=\mu-\Delta D$ with $\Delta D=|\mu|$}, at $t=40$ opens this channel. Unit 3 then grows, jumps onto the large cycle at $t\approx105$ and stays there after the drive ends\acmk{. It} is still there at $t=3000$. Units 1 and 2 settle at a small amplitude, $0.05$, driven by unit 3. The same increase moves every eigenvalue of the rest state away from the imaginary axis, and the slowest decay rate rises from $0.039$ to $0.051$. With the increase applied throughout a drive \hmk{of duration $300$}, the threshold falls as the drive grows (\hmk{$\Delta D/|\mu|=0.71$, $0.29$} and $0.14$ at \hmk{drive amplitudes} $0.5$, $1$ and $2$), and the same increase at unit 2 gives the same threshold.}

\subsection*{\hmk{Section S11. Opening in the nonlinear swing equation}}
\hmk{The traces of Fig.~\ref{fig:local-damping-fragility}A come from the linearized network. To test the opening beyond small disturbances, Eq.~\ref{eq:network_dynamics} with $g=\sin$ is integrated for the random ten-node network under the same noise at node \armk{5}, with a fourth-order Runge--Kutta scheme, a step of $10^{-3}$~s and $30$~s of record after a transient of $5$~s. The operating line angles $\theta^{\mathrm{op}}_n-\theta^{\mathrm{op}}_i$\armk{, set by a dispatch of $+1$ at node 4 and $-1$ at node 10 so that $\ell_{4,10}$ carries the largest of them,} are scaled so that the largest is $0^\circ$, $30^\circ$ or $45^\circ$\acmk{. The} couplings are $K_{in}=-L_{in}/\cos(\theta^{\mathrm{op}}_n-\theta^{\mathrm{op}}_i)$, so that the linearization reproduces $L$, and the injections $\Omega_i$ hold this operating point. The amplitude of the noise is raised until the largest excursion of a line angle from its operating value reaches $44^\circ$. Doubling the damping of node \armk{4} raises the rms flow on \armk{$\ell_{4,10}$} by a factor of \armk{$11.1$} in the linear limit\anmk{. With an operating line angle of $0^\circ$ the factor stays between \armk{$11.1$ and $11.2$} up to the largest excursion. With $30^\circ$ and $45^\circ$ it falls steadily as the excursion grows, to \armk{$6.2$ and $4.5$} at an excursion of \armk{$44^\circ$, and it falls by $25\%$ at an excursion of $25^\circ$ and $16^\circ$}}.\armk{ With the dispatch $P_i=(-1)^i$, which leaves $\ell_{4,10}$ nearly unloaded, the factor stays within $2\%$ of its linear value up to $44^\circ$ at all three operating angles.}}

\subsection*{\jmk{Section S12. Local changes of inertia}\aymk{ and line coupling}}
\jmk{\lmk{This section tests whether a change of inertia\aymk{, or of the coupling of a line,} opens latent channels as damping does. The inertia of a single node is lowered, with damping, couplings and the operating point unchanged.} A change $\Delta M_i$ of the inertia of node $i$ adds the rank-one term $-\omega^2\Delta M_i\,e_ie_i^\top$ to the dynamic stiffness matrix, so Eq.~\ref{eq:mobius_closed_form} holds with $j\omega\Delta D_i$ replaced by $-\omega^2\Delta M_i$\acmk{. In} the random ten-node network it agrees with a direct re-solution to within $10^{-11}$. Because $D\succ0$, $\mathrm{Im}\,\alpha_i<0$ at every $\omega>0$, so the denominator does not vanish. The increment is real, however, so the gain is not confined to \aemk{the disk whose diameter joins $G^{0}_{r,\ell}$ to the limit of Eq.~\ref{eq:ceiling}}, and the \acmk{ceiling} of the main text does not apply.}

\begin{table}[!htbp]
\centering
\caption{\textbf{Channel openings in the random networks under local changes of damping and inertia.} \acmk{Networks gives the share of networks, in percent, in which a change at a single node opens some channel by more than a factor of $2$ / $10$. Peak gives the largest opened peak among channels at less than a tenth of the strongest baseline flow under the same disturbance, in percent of that flow. Above gives the number of channels at less than half of the strongest flow that rise above it\bemk{, excluding changes at the disturbed node}. There are $300$ networks per size. Damping is the network's mean damping added at one node, and inertia is the inertia of one node halved.}}
\label{tab:inertia_ens}
\vspace{0.3em}
\begin{tabular}{lcccccc}
\hline
 & \multicolumn{3}{c}{Damping} & \multicolumn{3}{c}{Inertia $\times0.5$} \\
Size $N$ & Networks & Peak & Above & Networks & Peak & Above \\
\hline
$10$ & $23$ / $3$ & $19$ & $0$ & $99$ / $21$ & $83$ & $2$ \\
$20$ & $32$ / $9$ & $24$ & $0$ & $100$ / $33$ & $87$ & $1$ \\
$40$ & $52$ / $15$ & $22$ & $0$ & $100$ / $61$ & $89$ & $6$ \\
$80$ & $82$ / $46$ & $21$ & $0$ & $100$ / $88$ & $110$ & $14$ \\
\hline
\end{tabular}
\end{table}

\lmk{In the random networks of Fig.~\ref{fig:ensemble}, halving the inertia of one node opens channels more often and \admk{by larger factors} than the added damping (table~\ref{tab:inertia_ens}), while the spectrum again \acmk{remains nearly unchanged} (over all $45{,}000$ halvings, the spectral margin moves toward the imaginary axis by at most $2.3\%$ and the smallest damping ratio falls by at most $3.4\%$). As for damping, the opening is carried by the residue, with a median residue factor of $5.6$ and a median pole factor of $1.00$ over the channels opened by more than a factor of $2$. The channels that halved inertia opens respond more strongly at baseline, with a median $\nu$ of $0.046$ against $0.0011$ for damping, and their opened peaks lie at higher frequencies, with a median of $4.6$~Hz against $1.4$~Hz, consistent with an increment that grows as $\omega^2$. In $56\%$ of them the opening is largest when the inertia of the disturbed node itself is halved, which never occurs for the added damping.}%

\lmk{The openings carried by inertia reach higher absolute levels. Channels at less than a tenth of the strongest baseline flow under the same disturbance reach \aomk{up to} $110\%$ of it \aomk{at $N=80$ (table~\ref{tab:inertia_ens})}, where a channel at $5.5\%$ of the strongest flow becomes the strongest line after an opening by a factor of $20$. Channels at less than half of the strongest flow rise above it \aomk{at every size}. Under the added damping, channels at less than a tenth of the strongest flow reach at most \aomk{$24\%$} of it, and no channel below $85\%$ of the strongest flow rises above it. The total branch-flow energy under white noise, summed over all disturbed nodes, is unchanged by halving the inertia of any node (\armk{$0.998$ to $1.001$} of its baseline value in the ten-node network) and falls by \armk{$7$ to $8\%$} under the added damping\acmk{. Inertia thus} redistributes the response among lines and frequencies while damping also removes it.}%

\jmk{\lmk{In} the random ten-node network, \armk{halving the inertia of node 10 raises the $\ell_{4,10}$ channel by a factor of $27$ at $3.92$~Hz}, a resonance other than the \armk{$3.37$}~Hz resonance \admk{that damping opens}. Under the white-noise disturbance of Fig.~\ref{fig:local-damping-fragility}A, halving the inertia of node \armk{10} raises the rms flow on \armk{$\ell_{4,10}$} by a factor of \armk{$57$}\lmk{,} against \armk{$11$} for doubling \armk{the damping of node 4}.}

\aymk{Lowering the coupling of a single line acts in the same way. In the random ten-node network, lowering the coupling between nodes 1 and 4 by $10\%$ raises the $\ell_{4,10}$ channel to $17$ times its baseline, to $26\%$ of the strongest flow, while the \bdmk{smallest} damping ratio rises slightly.}%

\jmk{\textbf{\aimk{Local changes of inertia on the pan-European grid.}} For the inertia panels of fig.~\ref{fig:real-grid}, the inertia of one bus is \lmk{\asmk{lowered by $10\%$}}, with damping and dispatch unchanged, and the same sources, buses and lines are evaluated, including changes at the source bus. The added oscillation of a line is its largest increase in peak gain over all sources and changed buses, relative to the strongest baseline line of the same source, without the exclusions of the census. \asmk{Lowering the inertia by $10\%$ raises $1557$ lines} by more than $10\%$ of the strongest line, against $13$ for tripling the damping, and \asmk{$297$} and $10$ of these have their opened peak below $2$~Hz. \asmk{The $297$ lines are not weak at baseline, with a median $\nu$ of $0.31$, and for $69\%$ of them the largest increase occurs with the inertia lowered at the disturbance source.} The case of fig.~\ref{fig:real-grid}\asmk{E} is a latent channel, \asmk{with $\nu=0.012$: lowering the inertia of M\"{u}hleberg by $10\%$ under a disturbance at the same bus raises the Etupes--Bourguignon line by a factor of $9.6$ at $5.3$~\azmk{Hz.}}}%

\asmk{\textbf{Local changes of line coupling on the pan-European grid.} The linearized coupling of one line is multiplied by $0.9$ or $0.5$ at the operating point of \bdmk{section S6}, and every line that shares an end bus with it is monitored for the $468$ sources over $0.05$--$10$~Hz, with the exact rank-one update of the receptance. Line angle differences at the operating point have a median of $0.7^\circ$ and a maximum of $22.7^\circ$, so re-solving the power flow would change the coupling of the altered line by at most $2.1\%$ for the factor $0.9$, and by less than $5\%$ for $98\%$ of the lines for the factor $0.5$. Peaks are taken inside the band, the screening is that of the census, and the altered line and its parallel circuits are not counted. Counting only these adjacent lines, lowering the coupling by $10\%$ raises $222$ lines by more than a tenth of the strongest line under the same disturbance, $12$ of them with the opened peak below $2$~Hz\aymk{.} \aymk{H}alving it raises $1114$ lines\aymk{, $268$ of them below $2$~Hz}. Among channels opened to more than $5\%$ of the strongest line, the largest opening ratio on the census grid is $8.0$ for the factor $0.9$ and \aumk{$23.5$} for $0.5$\aymk{.} \aymk{R}esolved on a finer grid around their peaks, they are $6.6$ (fig.~\ref{fig:real-grid}F, at $7.8$~Hz) and \aumk{$23.8$, at $2.9$~Hz. For the factor $0.5$, the second largest, $22.7$ on the finer grid, raises a line from $3\%$ to $73\%$ of the strongest line}. The adjacent lines give a lower bound. For $40$ altered lines, $38$ of them drawn at random, monitoring every line for the factor $0.9$ raises the number of lines with an added oscillation above $2\%$ of the strongest line from $22$ to $216$, and above $10\%$ from $1$ to $10$. Of the $12$ altered lines whose largest added oscillation exceeds $1\%$ of the strongest line, $9$ reach it on a line that does not share a bus with them, larger than on the adjacent lines by a median factor of $1.8$. For $41$ altered lines, including those of the largest openings, the spectral margin changes by less than $0.01\%$.}

\bdmk{All three changes open channels (fig.~\ref{fig:real-grid}K). The openings under lowered inertia are more pronounced than under added damping, and of the three changes only the added damping is passive.}%

\end{document}